\pdfoutput=1

\documentclass[11pt]{article}

\usepackage[]{ACL2023}

\usepackage{times}
\usepackage{latexsym}

\usepackage[T1]{fontenc}

\usepackage[utf8]{inputenc}

\usepackage{microtype}

\usepackage{inconsolata}

\usepackage{graphicx}
\usepackage{amsmath}
\usepackage{algorithm}
\usepackage{algorithmic}
\usepackage{multirow}
\usepackage{bm}
\usepackage{booktabs}
\definecolor{Red}{RGB}{202,12,22}

\title{Towards Imperceptible Document Manipulations against \\ Neural Ranking Models}

\author{Xuanang Chen$^{\rm 1,2}$
        \quad Ben He$^{\rm 1,2}$
        \quad Zheng Ye$^{\rm 3}$
        \quad Le Sun$^{\rm 2}$
        \quad Yingfei Sun$^{\rm 1}$ \\
        $^{\rm 1}$University of Chinese Academy of Sciences, Beijing, China \\
        $^{\rm 2}$Institute of Software, Chinese Academy of Sciences, Beijing, China \\
        $^{\rm 3}$South-Central University for Nationalities, Wuhan, China \\
        \texttt{chenxuanang19@mails.ucas.ac.cn, benhe@ucas.ac.cn}\\
        \texttt{yezheng@scuec.edu.cn, sunle@iscas.ac.cn, yfsun@ucas.ac.cn}}

\begin{document}
\maketitle
\begin{abstract}
Adversarial attacks have gained traction in order to identify potential vulnerabilities in neural ranking models (NRMs), but current attack methods often introduce grammatical errors, nonsensical expressions, or incoherent text fragments, which can be easily detected. 
Additionally, current methods rely heavily on the use of a well-imitated surrogate NRM to guarantee the attack effect, which makes them difficult to use in practice.
To address these issues, we propose a framework called Imperceptible DocumEnt Manipulation (IDEM) to produce adversarial documents that are less noticeable to both algorithms and humans. 
IDEM instructs a well-established generative language model, such as BART, to generate connection sentences without introducing easy-to-detect errors, and employs a separate position-wise merging strategy to balance relevance and coherence of the perturbed text.
Experimental results on the popular MS MARCO benchmark demonstrate that IDEM can outperform strong baselines while preserving fluency and correctness of the target documents as evidenced by automatic and human evaluations.
Furthermore, the separation of adversarial text generation from the surrogate NRM makes IDEM more robust and less affected by the quality of the surrogate NRM.

\end{abstract}

\section{Introduction}
Adversarial Information Retrieval (AIR) has been a topic of significant attention from the research community~\cite{DBLP:journals/sigir/DavisonNC06,DBLP:journals/cas/LengKSM12,DBLP:journals/corr/abs-1911-11060}.
It refers to the scenario in which a portion of the collection can be maliciously manipulated, with Adversarial Web Search~\cite{DBLP:journals/ftir/CastilloD10} being a typical example.
In the context of the Web, this type of manipulation is commonly referred to as black-hat search engine optimization (SEO) or Web spamming, whose goal is to deceive ranking algorithms by artificially inflating the relevance of targeted Web pages, resulting in undeservedly high rankings for those pages~\cite{DBLP:conf/airweb/GyongyiG05}. 
Meanwhile, in the last few years, neural ranking models (NRMs), particularly those that utilize pre-trained language models (PLMs), have demonstrated remarkable performance across a diverse range of text ranking tasks~\cite{DBLP:series/synthesis/2021LinNY}.
Furthermore, these NRMs have also been implemented in various industrial applications~\cite{DBLP:series/synthesis/2021LinNY}, such as Web search engines~\cite{DBLP:conf/kdd/ZouZCMCWSCY21}, to improve the accuracy and relevance of search results.

\begin{figure}
    \centering
    \includegraphics[width=\linewidth]{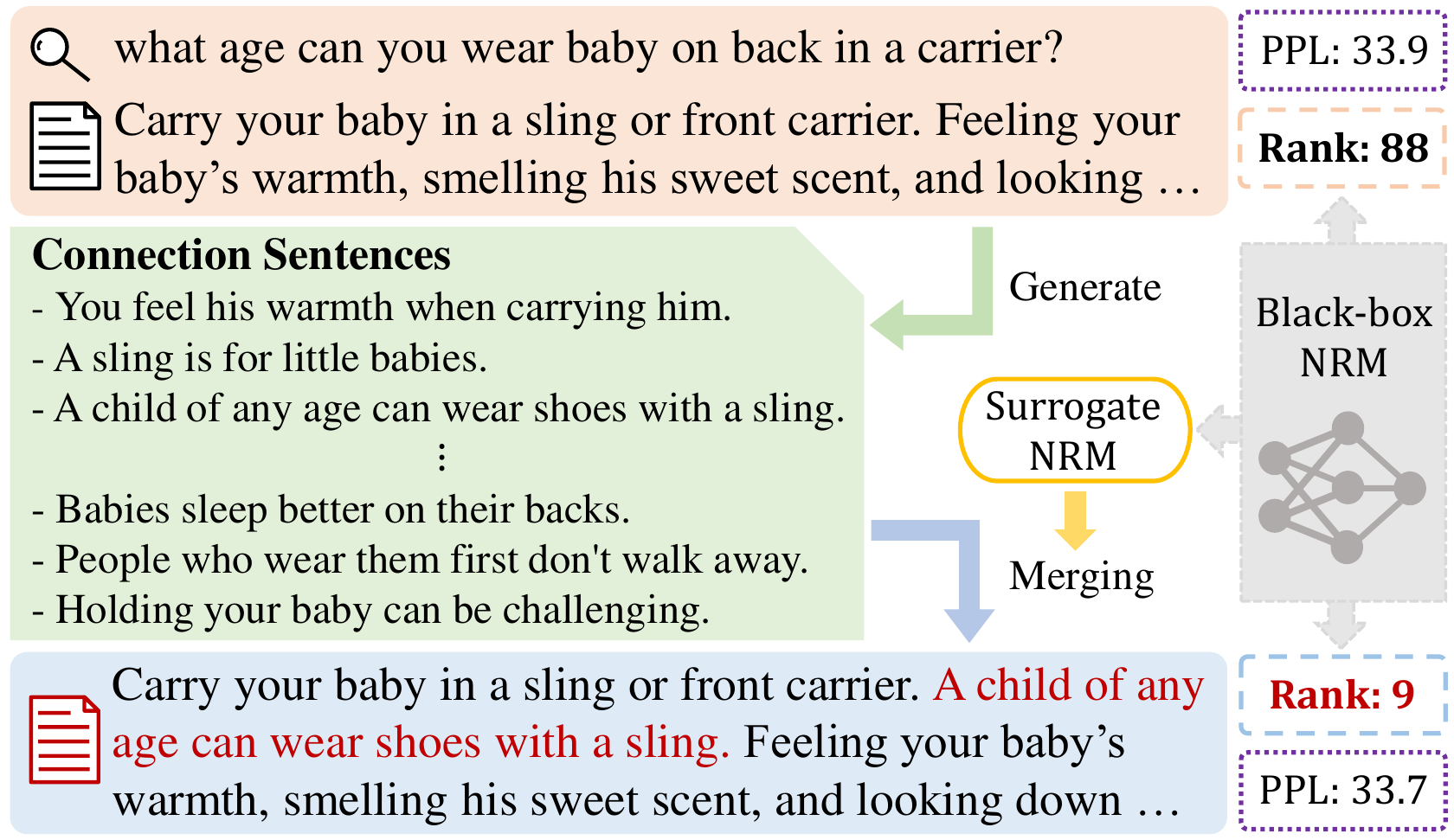}
    \caption{In IDEM, a generative language model is instructed to generate a group of connection sentences between the query and the target document, then a position-wise merging strategy is applied to select and position \textcolor{Red}{an optimal connection sentence} within the target document, in order to promote the ranking (from 88th to 9th) but maintain the fluency measured by perplexity (PPL).}
    \label{fig.idem}
\end{figure}

However, recent studies have revealed the vulnerabilities of NRMs to adversarial document manipulations~\cite{DBLP:journals/corr/abs-2204-01321,DBLP:conf/ccs/LiuKTSSWLL22,DBLP:conf/ictir/WangLA22,song-etal-2022-trattack}, that is, small deliberate perturbations in the input documents can cause a catastrophic ranking disorder in the outcome of NRMs.
This highlights the need to investigate and identify the potential weaknesses of NRMs before their deployment, in order to ensure their robustness and prevent potential risks.
Several manipulation techniques have been proposed to maliciously boost the ranking of low-ranked documents, such as PRADA~\cite{DBLP:journals/corr/abs-2204-01321} and PAT~\cite{DBLP:conf/ccs/LiuKTSSWLL22}. 
Although these adversarial attack methods have shown the ability to fool NRMs by replacing crucial words, or appending query text or other adversarial tokens, they are still subject to two major limitations.
Firstly, existing attack methods tend to introduce grammatical errors, impossible expressions, or incoherent text snippets into the original document, making the attack easy to mitigate, such as by perplexity filters~\cite{song-etal-2020-adversarial} or grammar checkers~\cite{DBLP:conf/ccs/LiuKTSSWLL22}.
Secondly, existing attack methods heavily rely on a well-imitated surrogate NRM to produce adversarial text, but this requires a lot of in-domain training data collected by querying the victim NRM, which can be infeasible or even unavailable in real-world situations.

To this end, we propose IDEM, an imperceptible document manipulation framework that aims to produce adversarial documents that are less perceptible to both humans and algorithms. 
As depicted in Figure~\ref{fig.idem}, a well-established generative language model (GLM), such as BART, is first engineered to generate a series of grammatically correct connection sentences between the query and the target document, a position-wise merging mechanism is then employed to select an optimal connection sentence to be appropriately positioned within the target document. 
During the generation of connection sentences, we take advantage of the language modeling objective in the GLM without adding any ranking-incentivized objective. 
This not only helps to produce more natural and fluent text without introducing extra errors that are easy to detect, but also separates the surrogate NRM from the generation process to substantially reduce the dependence of attack performance on the surrogate NRM. 
Extensive experiments carried out on the widely-used MS MARCO passage ranking dataset indicate that IDEM is able to achieve better attack performance against black-box NRMs than recent baselines, regardless of whether the surrogate NRM is similar to or far from the victim NRM. 
According to both automatic and human evaluations, the adversarial documents produced by IDEM can maintain semantic fluency and text quality.

Our contributions are three-fold: 1) We propose an imperceptible document manipulation framework, IDEM\footnote{We will make our code and data publicly available in the final version for future research.}, that employs contextualized blank-infilling and coherent merging for the ranking attack. 2) Extensive attack experiments under three types of surrogate NRMs show that IDEM is able to robustly promote the rankings of the target documents. 3) Automatic and human evaluations of text quality indicate that IDEM is capable of generating more natural and fluent adversarial documents.
\section{Problem Statement}\label{sec.problem}
\textbf{Task description.} 
Typically, given a query $q$ and candidates $\mathcal{D}\!=\!\{d_{1}, d_{2}, \cdots, d_{|\mathcal{D}|}\}$ collected by a lightweight retrieval model (e.g., BM25;~\citealp{DBLP:conf/trec/RobertsonWHGP95}) from the whole corpus, a NRM $\mathcal{M}_{V}$ (i.e., the victim NRM) produces relevance scores $s(q, d_{i})$ for all query-document pairs, outputting a re-ranking list $\mathcal{L}\!=\!\lbrack d_{1}, d_{2}, \cdots, d_{|\mathcal{D}|} \rbrack$, wherein $s(q, d_{1})\!>\!s(q, d_{2})\!>\!\cdots\!>\!s(q, d_{|\mathcal{D}|})$.
In the adversarial ranking attack, a deliberate perturbation $p_{i}$ is added into the target document $d_{i}\!\in\!\mathcal{L}$ to be an adversarial one $d_{i}^{adv}\!:=\!d_{i} \oplus p_{i}$, which can be ranked higher by the victim NRM.
Existing attack methods search and find perturbations on different levels of granularity, such as replacing important words with synonyms (e.g., PRADA) or adding an extra text piece (e.g., PAT).
Ideally, adversarial documents should display semantic consistency and fluency to avoid detection.
However, there is still room for improvement regarding the imperceptibility of adversarial examples produced by existing methods, such examples can be found in Appendix~\ref{sec.app.examples}.

\textbf{Task setting.}
For consistency with real-world situations (e.g., black-hat SEO), akin to recent studies~\cite{DBLP:journals/corr/abs-2204-01321,DBLP:conf/ccs/LiuKTSSWLL22}, this work focuses on the decision-based black-box attack setting, where attackers can only obtain a limited number of queries $\{q_{i}\}_{i=1}^{m}$ and their corresponding ranking lists $\{\mathcal{L}_{i}\!=\!\lbrack d_{i,1}, d_{i,2}, \cdots, d_{i,|\mathcal{D}|} \rbrack\}_{i=1}^{m}$ with rank positions by querying the victim NRM $\mathcal{M}_{V}$, but have no access to the exact relevance scores, as well as the architecture, parameters, gradients, or training data of the victim NRM.
In this setting, a weak-supervised training data $\mathcal{T}_{S}\!=\!\{(q_{i}, d_{i,j}, d_{i,k})\}_{i=1}^{m}$ is usually collected on basis of $\{(q_{i}, \mathcal{L}_{i})\}_{i=1}^{m}$, wherein $1\!\leq\!j\!<\!k\!\leq\!|\mathcal{D}|$ so that $d_{i,j}$ and $d_{i,k}$ are considered as relatively positive and negative documents, respectively. 
A surrogate NRM $\mathcal{M}_{S}$ is trained on this $\mathcal{T}_{S}$ to imitate the victim NRM $\mathcal{M}_{V}$ using a pair-wise loss, such as hinge loss~\cite{DBLP:journals/corr/abs-2204-01321}.
Besides, existing methods require this surrogate NRM to be functionally similar to the victim NRM, as it determines the direction of adversarial attack~\cite{DBLP:conf/ccs/LiuKTSSWLL22}.
This means the attack performance of existing methods heavily depends on the surrogate NRM, but it needs lots of in-domain training samples (i.e., $m$ is large) as well as the associated cost of querying the victim NRM, while our proposed IDEM can achieve reliable attack effect even with an out-of-domain (OOD) surrogate NRM.
 
\section{Method}\label{sec.method}
As outlined in Figure~\ref{fig.idem} and Algorithm~\ref{alg.idem}, IDEM contains two stages, the first is to obtain a series of connection sentences that can bridge the semantic gap between the query and the target document.
After that, these connection sentences are considered at all positions in the inter-sentences of the target document to trade-off the semantic fluency and attacked relevance, and the best one is output as the final adversarial document.

\subsection{Connection Sentences Generation}
To obtain more natural and fluent connection sentences, we choose to draw support from the well-established GLMs, such as BART in our default setting.
The BART model~\cite{lewis-etal-2020-bart} has been pre-trained using a text infilling loss function, which enables it to fill in blanks within the context in a more flexible way. 
Hence, in this section, we use the BART model as an example to illustrate how to engineer GLMs in a blank-infilling pattern to generate connection sentences that include information about both the query and the document.

Specifically, given a query $q$ and a target document $d_{i}\!\in\!\mathcal{L}$, which is simplified as $d$ from here, we concatenate them with a blank in between as seen in the template as follows.
\begin{equation}
    q \oplus {\rm It \, is \, known \, that \, \underline{\qquad} } \oplus d
    \label{eq.template}
\end{equation}
\noindent wherein query $q$ ends with appropriate punctuation, and the blank is replaced by a single special token defined and used in the GLM, such as \texttt{[MASK]} token in the BART model.
Given both query $q$ and document $d$ as the context, the prefix text ``It is known that'' serves as a prompt~\cite{DBLP:journals/corr/abs-2107-13586} to instruct the BART model to output more informative text that is related to both $q$ and $d$. 
The choice of prompt words can have an impact on the final attack effect, which is analyzed in Section~\ref{sec.exp.analysis}.

Afterward, the BART model takes Eq.~\ref{eq.template} as the input and fills a variable number of tokens (i.e., sentences with varying lengths) into the blank position.
Herein, we do not incorporate extra ranking-incentivized objectives into the BART model in order to produce grammatically correct connection sentences well suited to the surrounding text.
Besides, we employ the Top-\textit{k} sampling~\cite{fan-etal-2018-hierarchical} strategy to ensure the diversity of the generated sentences.
As depicted in Algorithm~\ref{alg.idem}, we sample multiple times (i.e., $K$), take $M$ candidates each time and save at most $N$ connection sentences, denoted as $\mathcal{S}_{c} = \{\overline{s}_{1}, \overline{s}_{2}, \cdots \overline{s}_{N}\}$.
We also limit the maximum length of the connection sentence to $L$ words to control the interference in the semantic content of the original document. 
All connection sentences in $\mathcal{S}_{c}$ are considered in the next stage for the merging with the original document $d$.

\begin{algorithm}[tb]
\caption{IDEM}
\label{alg.idem}
\textbf{Input}: a query $q$, a target document $d$, a surrogate NRM $\mathcal{M}_{S}$ for a victim NRM $\mathcal{M}_{V}$ \\
\textbf{Parameter}: the number of sentence in the target document $|d|$, the max number of connection sentence $N$, sample times $K$, sample size $M$, the max word length of connection sentence $L$ \\
\textbf{Output}: an adversarial document $d^{adv}$

\begin{algorithmic}[1] 
\STATE \textbf{Stage 1: Connection Sentences Generation}
\STATE Let $\mathcal{S}_{c}=\{\}$
\FOR{1 to $K$}
\WHILE{$|\mathcal{S}_{c}| < N$}
\STATE Sample $M$ connection sentence candidates $\{\overline{s}_{u}\}_{u=1}^{M}$ from the BART model by taking Eq.~\ref{eq.template} as the input
\IF {the length of $\overline{s}_{u}$ is smaller than $L$}
\STATE Keep $\overline{s}_{u}$ in $\mathcal{S}_{c}$
\ENDIF
\ENDWHILE
\ENDFOR
\STATE \textbf{Stage 2: Merging with Original Document}
\FOR{$\overline{s}_{u}$ in $\mathcal{S}_{c}$}
\FOR{position index $v$ from 0 to $|d|$}
\STATE Get an adversarial candidate $d^{adv}_{\langle u,v \rangle}$ as in Eq.~\ref{eq.adv.doc}, evaluate its \textit{coherence} score (Eq.~\ref{eq.nsp}) and \textit{relevance} score (Eq.~\ref{eq.rel}), and compute the weighted merging score (Eq.~\ref{eq.final})
\ENDFOR
\ENDFOR
\STATE Find and save the top-1 ranked $d^{adv}_{\langle u,v \rangle}$ as $d^{adv}$
\STATE \textbf{return} $d^{adv}$
\end{algorithmic}
\end{algorithm}

\subsection{Merging with Original Document}\label{sec.method.merge}
As Transformer-based NRMs have a positional bias towards the start of document~\cite{jiang-etal-2021-bert,DBLP:conf/ecir/HofstatterLAZH21}, adding adversarial text pieces at the beginning can usually achieve remarkable attack effect~\cite{DBLP:conf/ictir/WangLA22}, but this also has the obvious drawback that the attack can be easily detected~\cite{DBLP:conf/ccs/LiuKTSSWLL22}.
Thus, to balance between the attack effect and the fidelity of perturbed document content, we design a position-wise merging strategy to place an appropriate connection sentence at an optimal position within the original target document.

Specifically, given a query $q$, a target document with multiple sentences $d\!=\!s_{1}, s_{2}, \cdots\!, s_{|d|}$ and a set of connection sentences $\mathcal{S}_{c}$, we evaluate a large number of candidate adversarial documents with respect to the various combinations of $d$ and $\mathcal{S}_{c}$.
For each connection sentence $\overline{s}_{u} \in \mathcal{S}_{c}$, we merge $\overline{s}_{u}$ with the target document $d$ by placing it at position index $v$ (i.e., an integer from 0 to $|d|$), obtaining adversarial candidates in the form of Eq.~\ref{eq.adv.doc}.
\begin{equation}
    d^{adv}_{\langle u,v \rangle} = \begin{cases}
    \overline{s}_{u} \oplus d &  v=0 \\
    d_{1 \rightarrow v} \oplus \overline{s}_{u} \oplus d_{(v+1) \rightarrow |d|} & 0 < v < |d| \\
    d \oplus \overline{s}_{u} & v=|d| 
    \label{eq.adv.doc}
    \end{cases}
\end{equation}
wherein $d_{a \rightarrow b}$ means the text piece consists of consecutive sentences from $s_{a}$ to $s_{b}$, 
and the subscript $\langle u,v \rangle$ means the connection sentence $\overline{s}_{u}$ is inserted at the index $v$ of the target document $d$.

Subsequently, we examine all of the candidate adversarial documents $d^{adv}_{\langle u,v \rangle}$ and find out the best one as the final adversarial document $d^{adv}$.
An effective adversarial example should be imperceptible to human judges yet misleading to NRMs~\cite{DBLP:journals/corr/abs-2204-01321}. 
If the semantics of the added connection sentence differs greatly from the surrounding text in the target document, the resulting content will be incoherent and easily noticed by humans.
Thus, with the help of the next sentence prediction (NSP) function in the pre-trained BERT model~\cite{devlin-etal-2019-bert}, we define a coherence score on the junction between the connection sentence and the original target document.

Specifically, when the connection sentence $\overline{s}_{u}$ is placed in the middle of the target document $d$ (i.e., $0\!<\!v\!<\!|d|$), both the text pieces before and after the connection sentence are fed into the NSP function to get the \textit{coherence} score as seen in Eq.~\ref{eq.nsp}.
\begin{equation}
\begin{aligned}
    \textit{coh}(d^{adv}_{\langle u,v \rangle}) = \; & 0.5 \times \lbrack f_{\textit{nsp}}(d_{1 \rightarrow v}, \overline{s}_{u} \oplus d_{(v+1) \rightarrow |d|}) \\ 
    & + f_{\textit{nsp}}(d_{1 \rightarrow v} \oplus \overline{s}_{u}, d_{(v+1) \rightarrow |d|}) \rbrack
\label{eq.nsp}
\end{aligned}
\end{equation}
\noindent wherein $f_{\textit{nsp}}({\rm A,B})$ is the NSP score of text A being followed by text B. Similarly, when $\overline{s}_{u}$ is attached to the beginning or the end of the document $d$ (i.e., $v\!=\!0$ or $v\!=\!|d|$), the coherence score is given as $f_{\textit{nsp}}(\overline{s}_{u}, d)$ or $f_{\textit{nsp}}(d, \overline{s}_{u})$, respectively.

Apart from the semantic coherence, another important factor that impacts the feasibility of attack is the relevance between the query $q$ and the candidate $d^{adv}_{\langle u,v \rangle}$.
Herein, due to the black-box setting as described in Section~\ref{sec.problem}, we need a surrogate NRM $\mathcal{M}_{S}$ to estimate the \textit{relevance} score as in Eq.~\ref{eq.rel}.
\begin{equation}
    \textit{rel}(q, d^{adv}_{\langle u,v \rangle}) = \mathcal{M}_{S}(q, d^{adv}_{\langle u,v \rangle})
    \label{eq.rel}
\end{equation}
\noindent wherein the surrogate NRM $\mathcal{M}_{S}$ is not required to be heavily ranking-consistent with the victim NRM $\mathcal{M}_{V}$ but is only expected to distinguish which connection sentence contains spurious-relevant information (e.g., query keywords). 
In other words, the surrogate NRM does not participate in the generation of connection sentences, so that the adversarial information is from the BART model rather than the surrogate NRM.
Thus, IDEM does not compulsively require a well-imitated surrogate NRM that is trained on a large number of in-domain training samples.
As demonstrated later in Section~\ref{sec.exp.res}, even when using an OOD surrogate NRM that is only slightly better than BM25, our proposed IDEM still achieves relatively high attack performance.

Finally, to trade-off the semantic coherence and relevance, as seen in Eq.\ref{eq.final}, we resort to a weighted sum of them as the merging score for each $d^{adv}_{\langle u,v \rangle}$.
\begin{equation}
\begin{aligned}
    \textit{score}_{m}(q, d^{adv}_{\langle u,v \rangle}) = \; & \alpha \times \textit{coh}(d^{adv}_{\langle u,v \rangle}) \\
    & + (1-\alpha) \times \textit{rel}(q, d^{adv}_{\langle u,v \rangle}) 
    \label{eq.final}
\end{aligned}
\end{equation}
\noindent wherein $\alpha$ is a weight factor, and both scores are transformed to $[0, 1]$ by the min-max normalization before being added together.
Out of all the candidates $d^{adv}_{\langle u,v \rangle}$, the one with the highest merging score is chosen as the final adversarial document $d^{adv}$. 
\section{Experiments}
\subsection{Experimental Setup}\label{sec.exp.setup}
\textbf{Dataset.} 
Akin to previous works~\cite{DBLP:journals/corr/abs-2204-01321,DBLP:conf/ictir/WangLA22,DBLP:conf/ccs/LiuKTSSWLL22}, we employ popular MS MARCO passage dataset~\cite{DBLP:conf/nips/NguyenRSGTMD16} for our experiments, which contains about 8.8M passages (seen as documents in this paper) as the corpus. The evaluation data contains a Dev set with 6,980 queries and an Eval set with 6,837 queries.
In addition, we use Natural Question (NQ) dataset~\cite{DBLP:journals/tacl/KwiatkowskiPRCP19} that has been processed by~\citet{karpukhin-etal-2020-dense} as the OOD training source for the surrogate NRM.

\textbf{Victim NRM.}
Akin to~\citet{DBLP:conf/ccs/LiuKTSSWLL22}, the `ms-marco-MiniLM-L-12-v2' model publicly available at Sentence-Transformers~\cite{reimers-gurevych-2019-sentence} is used as the representative victim NRM (i.e., $\mathcal{M}_{V}$) in our study. 
This victim NRM adopts MiniLM~\cite{DBLP:conf/nips/WangW0B0020} as the backbone, and it is fine-tuned on MS MARCO in a cross-encoder architecture~\cite{DBLP:journals/corr/abs-1901-04085}. 
Overall, $\mathcal{M}_{V}$ achieves highly effective ranking performance on the MS MARCO Dev set as seen in Table~\ref{tab:surrogate.res}.

\textbf{Surrogate NRMs.}
To examine the practicality of ranking attack methods in real-world situations, wherein access to the victim ranking system could be limited or unavailable, we experiment with three kinds of surrogate NRMs: 
\bm{$\mathcal{M}_{S_1}$} is trained on all 6,837 Eval queries, \bm{$\mathcal{M}_{S_2}$} is trained on the randomly sampled 200 Eval queries, and \bm{$\mathcal{M}_{S_3}$} is trained on the OOD NQ queries.
More details of surrogate NRMs can be found in Appendix~\ref{sec.app.surro}.

\textbf{Target documents.}
Following~\citet{DBLP:journals/corr/abs-2204-01321}, we evaluate attack methods on randomly sampled 1K Dev queries with two types of target documents in different attack difficulties, i.e., Easy-5 and Hard-5, which are sampled from the re-ranked results by the victim NRM on the top-1K BM25 candidates.
Specifically, \textbf{Hard-5} is the five bottom-ranked documents, and \textbf{Easy-5} is the five documents ranked between $\lbrack51, 100\rbrack$ in which one document is randomly picked out of every 10 documents.

\textbf{Details of IDEM.}
By default, pre-trained BART-Base~\cite{lewis-etal-2020-bart} is instructed to generate connection sentences.
The impacts of other GLMs, including pre-trained T5~\cite{DBLP:journals/jmlr/RaffelSRLNMZLL20} and fine-tuned GPT-2~\cite{donahue-etal-2020-enabling}, are also examined in Section~\ref{sec.exp.analysis}.
In the generation stage, $\textit{k}$ in sampling strategy is set as 50, and we sample 10 or 50 times (50 ones per time) and save at most 100 or 500 connection sentences with a max length of 12 words (i.e., $M\!=\!50$, $K\!=\!10/50$, $N\!=\!100/500$ and $L\!=\!12$).
In the merging stage, we employ the NSP function in pre-trained BERT-Base to evaluate the coherence, and the $\alpha$ in Eq.\ref{eq.final} is set as 0.5 (0.1) for Easy-5 (Hard-5) target documents. 
The impact of $\alpha$ on the attack results is shown in Appendix~\ref{sec.app.alpha}.

\begin{table*}[t]
\centering
\resizebox{\textwidth}{!}{
\begin{tabular}{c|l|ccccc|ccccc}
\toprule
  \textbf{Surrogate} & \multirow{2}{*}{\textbf{Method}} & \multicolumn{5}{c|}{\textbf{Easy-5}} & \multicolumn{5}{c}{\textbf{Hard-5}} \\
  \textbf{NRM} & & \textbf{ASR} & \bm{$\% \, r\!\leq\!10$} & \bm{$\% \, r\!\leq\!50$} & \textbf{\textit{Boost}} & \textbf{PPL}$\downarrow$ & \textbf{ASR} & \bm{$\% \, r\!\leq\!10$} & \bm{$\% \, r\!\leq\!50$} & \textbf{\textit{Boost}} & \textbf{PPL}$\downarrow$ \\
  \midrule
  - & Original & - & - & - & - & 37.3 & - & - & - & - & 50.5 \\ 
  - & Query+ & 100.0 & 86.9 & 99.2 & 70.3 & 45.4 & 100.0 & 47.8 & 78.3 & 955.1 & 67.5 \\ 
  \midrule
  \multirow{5}{*}{$\mathcal{M}_{S_{1}}$} & PRADA & 77.9 & 3.52 & 46.2 & 23.2 & 94.4 & 68.0 & 0.02 & 0.10 & 65.2 & 154.4 \\
  & Brittle-BERT & 98.7 & 81.3 & 96.7 & 67.3 & 107.9 & \textbf{100.0} & \textbf{61.5} & \textbf{85.9} & \textbf{965.5} & 152.5 \\
  & PAT & 89.6 & 30.6 & 73.8 & 41.9 & 50.9 & 98.0 & 6.24 & 20.1 & 589.1 & 71.4  \\
  & IDEM$_{N = 100}$ & 99.3 & 77.9 & 97.0 & 67.0 & \textbf{36.2} & 99.5 & 41.4 & 66.9 & 875.6 & \textbf{54.6} \\
  & IDEM$_{N = 500}$ & \textbf{99.7} & \textbf{87.4} & \textbf{99.0} & \textbf{70.3} & 36.4 & 99.8 & 54.3 & 79.3 & 933.0 & 54.9 \\
  \midrule
  \multirow{5}{*}{$\mathcal{M}_{S_{2}}$} & PRADA & 69.6 & 1.36 & 35.0 & 17.4 & 90.7 & 66.0 & 0.00 & 0.10 & 49.3 & 152.1 \\
  & Brittle-BERT & 81.6 & 33.2 & 69.7 & 36.4 & 131.3 & 94.8 & 8.98 & 25.4 & 565.1 & 179.5 \\
  & PAT & 61.9 & 8.46 & 37.3 & 12.5 & 49.3 & 84.4 & 0.82 & 3.40 & 221.7 & 66.2 \\
  & IDEM$_{N = 100}$ & 96.8 & 65.3 & 91.8 & 60.7 & \textbf{36.5}& 97.6 & 31.7 & 57.0 & 822.0 & \textbf{54.7} \\
  & IDEM$_{N = 500}$ & \textbf{98.7} & \textbf{74.8} & \textbf{95.4} & \textbf{65.1} & 37.0 & \textbf{99.1} & \textbf{39.6} & \textbf{67.4} & \textbf{890.7} & 55.4 \\
  \midrule
  \multirow{5}{*}{$\mathcal{M}_{S_{3}}$} & PRADA & 71.5 & 1.86 & 37.5 & 19.1 & 91.5 & 71.9 & 0.00 & 0.08 & 73.4 & 168.7 \\
  & Brittle-BERT & 90.0 & 43.4 & 80.1 & 46.2 & 117.7 & \textbf{99.9} & 17.7 & 47.6 & 845.2 & 156.8 \\
  & PAT & 51.1 & 2.70 & 22.9 & 2.01 & 46.8 & 79.0 & 0.00 & 0.66 & 92.9 & 64.2 \\
  & IDEM$_{N = 100}$ & 97.2 & 57.3 & 89.8 & 58.1 & \textbf{36.9} & 99.2 & 23.0 & 48.8 & 805.7 & \textbf{55.2} \\
  & IDEM$_{N = 500}$ & \textbf{98.8} & \textbf{65.3} & \textbf{93.8} & \textbf{61.9} & 37.7 & 99.8 & \textbf{29.1} &  \textbf{57.9} & \textbf{866.2} & 56.0 \\
  \bottomrule
\end{tabular}
}
\caption{\label{tab:attack.res}
The attack results on two types of target documents under three kinds of surrogate NRMs for sampled 1K queries from the MS MARCO Dev set. A lower PPL is better, while other metrics are the higher the better.}
\end{table*}

\begin{table}[t]
\centering
\resizebox{0.99\linewidth}{!}{
\begin{tabular}{l|cccc}
\toprule
  \textbf{Model} & \textbf{MRR@10} & \textbf{MRR@1K} & \textbf{Inter@10} & \textbf{RBO@1K} \\
  \midrule
  BM25 & 18.4 & 19.5 & 22.8 & 31.5 \\
  \midrule
  $\mathcal{M}_{V}$ & 39.5 & 40.3 & - & - \\
  \midrule
  $\mathcal{M}_{S_1}$ & 37.0 & 37.8 & 73.1 & 66.2 \\
  $\mathcal{M}_{S_2}$ & 23.0 & 24.2 & 41.0 & 31.2 \\
  $\mathcal{M}_{S_3}$ & 21.0 & 22.2 & 27.3 & 37.6 \\
\bottomrule
\end{tabular}
}
\caption{The re-ranking performance of the victim and surrogate NRMs on the MS MARCO Dev set (BM25 Top-1K). 
Inter@10 and RBO@1K indicate the ranking consistency relative to the victim NRM $\mathcal{M}_{V}$.}
\label{tab:surrogate.res}
\end{table}

\textbf{Compared methods.}
We compare the following attack methods against NRMs: 
\textbf{Query+}~\cite{DBLP:conf/ccs/LiuKTSSWLL22} directly adds the query text at the beginning of the target document. 
\textbf{PRADA}~\cite{DBLP:journals/corr/abs-2204-01321} finds and replaces important words with synonyms in the target document.
\textbf{Brittle-BERT}~\cite{DBLP:conf/ictir/WangLA22} and \textbf{PAT}~\cite{DBLP:conf/ccs/LiuKTSSWLL22} append a few tokens at the beginning of the target document, the former only considers the ranking-oriented objective, while the later further considers semantic and fluency constraints.
For fair comparisons, PRADA perturbs at most 20 tokens, Brittle-BERT and PAT add at most 12 tokens.
More details of these attack baselines are available in Appendix~\ref{sec.app.baseline}.

\textbf{Metrics.}
As for the re-ranking results, we report official MRR@10/1K metrics on the MS MARCO Dev set. 
Akin to~\citet{DBLP:conf/ccs/LiuKTSSWLL22}, we calculate the overlap of top-10 (Inter@10) and the Rank Biased Overlap~\cite{DBLP:journals/tois/WebberMZ10} (RBO@1K, $p$ is set as 0.7) to measure the ranking consistency between the surrogate and victim NRMs.
As for the attack results, we report the percentage of successfully boosted target documents (i.e., attack successful rate, ASR) following~\citet{DBLP:journals/corr/abs-2204-01321}.
Akin to~\citet{DBLP:conf/ccs/LiuKTSSWLL22}, we also report the average boosted ranks (\textit{Boost}) of the target documents, and the percentage of the target documents that are promoted into top-10 ($\% \, r\!\leq\!10$) and top-50 ($\% \, r\!\leq\!50$).
In addition, we measure the average perplexity (PPL) calculated using a pre-trained GPT-2 model~\cite{radford2019language}, where a \textit{lower} PPL value reflects better fluency of the adversarial documents as suggested by~\citet{kann-etal-2018-sentence} and~\citet{lei-etal-2022-phrase}.

\subsection{Results}\label{sec.exp.res}
\textbf{IDEM demonstrates the ability to achieve comparable attack performance while not greatly diminishing the semantic fluency.}
The attack results are summarized in Table~\ref{tab:attack.res}, where all values are averaged across the 5K target documents for 1K Dev queries.
When the query is added to the beginning of the target document (referred to as Query+), a notable increase in ranking can be observed, which is as expected.
However, this approach negatively impacts semantic fluency, resulting in a loss of approximately 8 (17) perplexity points on Easy-5 (Hard-5) target documents.
Ideally, when the victim NRM is freely accessible, a sufficient amount of surrogate training data is obtainable to train a well-imitated surrogate model, like $\mathcal{M}_{S_1}$ in Table~\ref{tab:surrogate.res}.
As demonstrated in Table~\ref{tab:attack.res}, when applied in conjunction with $\mathcal{M}_{S_1}$, the use of word-level synonym substitutions (i.e., PRADA) does not provide  superior attack performance, but adding adversarial tokens at the beginning of the target documents (i.e., Brittle-BERT and PAT) yields better attack performance.
However, adversarial documents generated by these prior methods, particularly PRADA and Brittle-BERT, exhibit excessively high perplexity (PPL) values. 
In contrast, our proposed IDEM not only achieves comparable attack performance, exemplified by its superior performance on Easy-5 target documents, but also mitigates the detrimental effects on text fluency, as evidenced by the lower PPL values on its adversarial documents.

\textbf{IDEM is more robust and much less affected by the surrogate NRM, even when it is an OOD NRM.}
When access to the victim NRM (i.e., $\mathcal{M}_{V}$) is restricted, as shown in Table~\ref{tab:surrogate.res}, there is a notable discrepancy between the surrogate $\mathcal{M}_{S_2}$ and the victim $\mathcal{M}_{V}$, as only a limited number of in-domain training samples can be used.
When working with $\mathcal{M}_{S_2}$, the attack performances of all methods, particularly PAT and Brittle-BERT, are greatly diminished as can be observed in Table~\ref{tab:attack.res}.
For example, when $\mathcal{M}_{S_1}$ is changed to $\mathcal{M}_{S_2}$, the attack performance ($\% \, r\leq10$) of Brittle-BERT drops dramatically from 81.3 to 33.2 on Easy-5 target documents, and from 61.5 to 8.98 on Hard-5 target documents.
Additionally, the semantic fluency of adversarial documents produced by Brittle-BERT also decreases, as evidenced by an increase in PPL from 107.9 to 131.3 on Easy-5 target documents. 
Furthermore, when access to the victim NRM is not available, only publicly available OOD training samples can be used, yield a more discrepant surrogate NRM, like $\mathcal{M}_{S_3}$ in Table~\ref{tab:surrogate.res}.
In this situation, as shown in Table~\ref{tab:attack.res}, the performances of prior attack methods are also not ideal, especially PRADA and PAT. 
However, when working with both $\mathcal{M}_{S_2}$ and $\mathcal{M}_{S_3}$, IDEM demonstrates the best attack results among all baselines while preserving the semantic fluency of target documents.
For more details, including the attack efficiency and adversarial examples for different attack methods, please refer to Appendixes~\ref{sec.app.time} and~\ref{sec.app.examples}, respectively.

\begin{table*}[t]
\centering
\resizebox{0.98\textwidth}{!}{
\begin{tabular}{l|ccccc|ccccc}
\toprule
  \multirow{2}{*}{\textbf{Prompt}} & \multicolumn{5}{c|}{\textbf{Easy-5}} & \multicolumn{5}{c}{\textbf{Hard-5}} \\
  & \textbf{ASR} & \bm{$\% \, r\!\leq\!10$} & \bm{$\% \, r\!\leq\!50$} & \textbf{\textit{Boost}} & \textbf{PPL}$\downarrow$ & \textbf{ASR} & \bm{$\% \, r\!\leq\!10$} & \bm{$\% \, r\!\leq\!50$} & \textbf{\textit{Boost}} & \textbf{PPL}$\downarrow$ \\
  \midrule

  \multicolumn{11}{c}{\textbf{IDEM$_{N = 100}$}} \\
  \midrule
  It is known that & 99.3 & 77.9 & 97.0 & 67.0 & 36.2 & 99.5 & 41.4 & 66.9 & 875.6 & 54.6 \\
  It is about that & 98.3 & 60.9 & 92.7 & 60.3 & 39.1 & 99.5 & 20.0 & 45.7 & 798.9 & 57.3 \\
  We know that & 99.3 & 74.4 & 96.7 & 65.8 & 36.8 & 99.4 & 33.9 & 59.8 & 851.1 & 54.9 \\
  The fact is that & 99.6 & 77.2 & 97.4 & 66.9 & 36.8 & 99.5 & 37.2 & 63.3 & 870.0 & 54.9 \\
  \midrule

  \multicolumn{11}{c}{\textbf{IDEM$_{N = 500}$}} \\
  \midrule
  It is known that & 99.7 & 87.4 & 99.0 & 70.3 & 36.4 & 99.8 & 54.3 & 79.3 & 933.0 & 54.9 \\
  It is about that & 99.6 & 75.5 & 97.6 & 66.7 & 39.4 & 99.9 & 33.5 & 63.7 & 892.6 & 58.4 \\
  We know that & 99.8 & 85.7 & 99.0 & 69.8 & 37.2 & 99.9 & 47.4 & 74.3 & 920.0 & 55.8 \\
  The fact is that & 99.8 & 87.0 & 99.1 & 70.3 & 36.7 & 99.9 & 50.1 & 76.5 & 928.1 & 55.5 \\
  
  \bottomrule
\end{tabular}
}
\caption{\label{tab:prompt.res}
The attack results of IDEM using different prompts to generate connection sentences. The surrogate NRM $\mathcal{M}_{S_1}$ is used. Overall, ``It is known that'' produces best attack results so that it is used by default in IDEM.}
\end{table*}

\begin{table}[t]
\centering
\resizebox{\linewidth}{!}{
\begin{tabular}{l|ccccc}
\toprule
  \multirow{2}{*}{\textbf{NRM}} & \multicolumn{5}{c}{\textbf{Index}} \\
  & \bm{$\%\,v\!=\!0$} & \bm{$\%\,v\!=\!1$} & \bm{$\%\,v\!=\!2$} & \bm{$\%\,v\!=\!3$} & \bm{$\%\,v\!\geq\!4$}  \\
  \midrule
  \multicolumn{6}{c}{\textbf{Easy-5}}\\
  \midrule
  $\mathcal{M}_{S_1}$ & 83.9 & 12.2 & 2.62 & 0.88 & 0.48 \\
  $\mathcal{M}_{S_2}$ & 74.1 & 15.0 & 4.77 & 1.56 & 4.59 \\
  $\mathcal{M}_{S_3}$ & 22.3 & 29.5 & 22.3 & 11.7 & 14.2 \\
  \midrule
 \multicolumn{6}{c}{\textbf{Hard-5}}\\
 \midrule
  $\mathcal{M}_{S_1}$ & 83.2 & 13.5 & 2.12 & 0.76 & 0.40 \\
  $\mathcal{M}_{S_2}$ & 76.3 & 14.6 & 4.32 & 1.12 & 3.68 \\
  $\mathcal{M}_{S_3}$ & 9.76 & 37.7 & 24.8 & 12.9 & 14.9 \\

  \bottomrule
\end{tabular}
}
\caption{Statistics on the positioning of connection sentences in adversarial documents created by IDEM$_{N\!=\!100}$.}
\label{tab:position.res}
\end{table}

\begin{table}[t]
\centering
\resizebox{0.95\linewidth}{!}{
\begin{tabular}{l|ccccc}
\toprule
  \textbf{GLM} & \textbf{ASR} & \bm{$\% \, r\!\leq\!10$} & \bm{$\% \, r\!\leq\!50$} & \textbf{\textit{Boost}} & \textbf{PPL}$\downarrow$ \\
  \midrule
  \multicolumn{6}{c}{\textbf{Easy-5}}\\
  \midrule
  BART & 99.3 & 77.9 & 97.0 & 67.0 & 36.2 \\
  T5 & 98.7& 75.3 & 95.4 & 65.2 & 38.0 \\
  ILM & 93.9 & 45.7 & 81.5 & 50.6 & 41.5 \\
  \midrule
 \multicolumn{6}{c}{\textbf{Hard-5}}\\
 \midrule
  BART & 99.5 & 41.4 & 66.9 & 875.6 & 54.6 \\
  T5 & 97.8 & 37.1 & 60.1 & 820.9 & 56.8 \\
  ILM & 98.3 & 16.0 & 35.6 & 693.2 & 59.8 \\

  \bottomrule
\end{tabular}
}
\caption{The attack results of IDEM$_{N\!=\!100}$ based on various GLMs and surrogate NRM $\mathcal{M}_{S_1}$.}
\label{tab:extend.res}
\end{table}

\subsection{Analysis}\label{sec.exp.analysis}
\textbf{Impact of prompt in IDEM.}
During the generation of connection sentence, a prompt (in Eq.\ref{eq.template}) is utilized to guide the BART model to output more informative text.
Herein, we examine the impact of prompt on the final attack results of IDEM.
As seen in Table~\ref{tab:prompt.res}, four different prompts are analyzed, in which ``It is known that'' produces better attack results, particularly on Hard-5 target documents, and also generates lower PPL values.
In contrast, ``The fact is that'' and ``We know that'' perform slightly worse on Hard-5 target documents, and ``It is about that'' performs even worse.
Overall, the choice of prompt has marginal impact on the ASR but mainly affects the degree of promotion in the attack.

\textbf{Position of connection sentence in IDEM.}
As mentioned in Section~\ref{sec.method.merge}, IDEM can automatically place an adversarial connection sentence within the target documents.
As summarized in Table~\ref{tab:position.res}, we count the positions (i.e., index $v$) of connection sentences.
When the surrogate NRM is $\mathcal{M}_{S_1}$ or $\mathcal{M}_{S_2}$, it is observed that less than 30\% of the connection sentences are positioned in the middle (i.e., $v\!\geq\!1$). 
Meanwhile, when the surrogate NRM is $\mathcal{M}_{S_3}$, the distribution of insert positions is found to be more uniform, with only 10-20\% of the connection sentences being appended to the beginning (i.e., $v\!=\!0$). 
These findings indicate that IDEM can place the adversarial text at any positions within the target document, making it harder to be detected.

\textbf{IDEM based on other GLMs.}
In addition to evaluating IDEM using BART, we also examine how well IDEM performs when other GLMs with blank-filling capabilities are instructed to generate connection sentences, including T5-Base~\cite{DBLP:journals/jmlr/RaffelSRLNMZLL20} and ILM (a fine-tuned version of GPT-2;~\citealp{donahue-etal-2020-enabling}).
As presented in Table~\ref{tab:extend.res}, our empirical results show that while the BART-based IDEM is found to be the overall best, all three models achieve high attack success rates, indicating the general applicability of the IDEM approach.

\textbf{Online grammar checking.}
In order to examine the extra errors introduced by different attack methods, we employ the popular online grammar checker \textit{Grammarly}\footnote{\url{https://app.grammarly.com}} to evaluate the quality of adversarial documents. 
Specifically, we collect 100 adversarial documents (with the same id) produced by each attack method for 50 Dev queries.
We report three evaluation metrics given by Grammarly, including the average number of issues in correctness (e.g., spelling, grammar, and punctuation) and suggestions (e.g., wordy or unclear sentences, etc.) in each adversarial document, and the quality score of all adversarial documents.
As shown in Table~\ref{tab:grammar.res}, IDEM introduces the fewest issues and obtains the highest quality score among all attack methods, indicating that the adversarial documents produced by IDEM are more machine-imperceptible.

\begin{table}[t]
\centering
\resizebox{\linewidth}{!}{
\begin{tabular}{l|ccc}
\toprule
  \textbf{Method} & \textbf{\#Correctness}$\downarrow$ & \textbf{\#Suggestions}$\downarrow$ & \textbf{Quality} \\
  \midrule
  Original & 1.80 & 4.23 & 59 \\
  \midrule
  Query+ & 2.39 & 6.50 & 49 \\
  PRADA & 5.51 & 11.0 & 22 \\
  Brittle-BERT & 3.74 & 7.12 & 41 \\
  PAT & 2.40 & 6.21 & 52 \\
  IDEM$_{N = 100}$ & 2.29 & 5.06 & 57 \\
  IDEM$_{N = 500}$ & 2.31 & 5.18 & 57 \\

\bottomrule
\end{tabular}
}
\caption{Automatic evaluation results provided by the online grammar checker (i.e., Grammarly).}
\label{tab:grammar.res}
\end{table}

\begin{table}[t]
\centering
\resizebox{0.75\linewidth}{!}{
\begin{tabular}{l|cc}
\toprule
  \multirow{2}{*}{\textbf{Method}} & \multicolumn{2}{c}{\textbf{Human-Imperceptibility}} \\
  & \textbf{\quad Avg. \quad} & \textbf{Kappa} \\
  \midrule
  Original & 0.69 & 0.44 \\
  \midrule
  Query+ & 0.36 & 0.55 \\
  PRADA & 0.45 & 0.56 \\
  Brittle-BERT & 0.50 & 0.43 \\
  PAT & 0.55 & 0.69 \\
  IDEM$_{N = 100}$ & 0.78 & 0.11 \\
  IDEM$_{N = 500}$ & 0.64 & 0.54 \\

\bottomrule
\end{tabular}
}
\caption{Human evaluation results on the imperceptibility of adversarial documents. 
The closer the average score is to 1, the less likely it is to be noticed by humans.}
\label{tab:human.res}
\end{table}

\textbf{Human evaluation.}
To further prove that the adversarial documents generated by IDEM are more natural to readers, a human-subject evaluation was conducted to assess the imperceptibility of adversarial text. 
Specifically, 32 adversarial documents were randomly selected for each attack method, and 32 original unaltered documents were also selected. 
All these documents were then mixed and randomly divided into two groups, with each group being evaluated by two annotators who are computer science graduate students with the necessary knowledge to understand the nature of the ranking attack. 
The annotators were tasked with determining whether the document content had been attacked (0) or was normal (1). 
We averaged all annotations on 32 samples from 2 annotators as the final imperceptibility score, and also computed Kappa coefficient for the annotation consistency.
As can be seen in Table~\ref{tab:human.res}, IDEM receives the highest score for human imperceptibility among all attack methods.
Additionally, the Kappa values of almost all attack methods are larger than 0.4 (considered as ``moderate agreement''), while IDEM$_{N=100}$ has the smallest Kappa value (i.e., 0.11), which seems reasonable since it is hard to reach an agreement on the more imperceptibly attacked documents.
\section{Related Work}
The adversarial IR has been studied for an extended period, such as black-hat SEO, which refers to the intentional manipulation of Web pages with the goal of obtaining an unjustified ranking position, resulting in a decline in the quality of search results and an inundation of irrelevant pages~\cite{DBLP:conf/airweb/GyongyiG05}. 
In this context, mainstream research focuses on studying the adversarial manipulations through various aspects, such as detection~\cite{DBLP:conf/kdd/DalviDMSV04,DBLP:conf/www/NtoulasNMF06}, theoretical and empirical analysis~\cite{DBLP:conf/sigir/RaiferRTK17}, robustness of LTR-based ranking functions~\cite{DBLP:conf/sigir/GorenKTR18}, automatic content modification~\cite{DBLP:conf/sigir/GorenKTR20}, and other research directions~\cite{DBLP:conf/sigir/KurlandT22}.

Recently, there has been a significant advancement in NRMs, particularly those based on PLMs (e.g., BERT), which have achieved exceptional text ranking performance~\cite{DBLP:series/synthesis/2021LinNY}.
At the same time, an increasing number of studies have highlighted the robustness concerns of NRMs in various settings, including the presence of query typos or variations~\cite{zhuang-zuccon-2021-dealing,DBLP:conf/ecir/PenhaCH22}, textual noises~\cite{DBLP:journals/ipm/ChenHHSS23}, and adversarial attacks~\cite{DBLP:journals/corr/abs-2008-02197,song-etal-2022-trattack}.
Although current attack methods, e.g., PRADA~\cite{DBLP:journals/corr/abs-2204-01321}, Brittle-BERT~\cite{DBLP:conf/ictir/WangLA22} and PAT~\cite{DBLP:conf/ccs/LiuKTSSWLL22}, can successfully mislead NRMs, they introduce additional quality issues and tend to heavily rely on the surrogate NRMs during the document manipulation.
Instead, our proposed IDEM effectively addresses the limitations of existing attack methods and demonstrates remarkable attack performance.

\section{Conclusion}
In this study, we introduce a document manipulation framework named IDEM, which is engineered to produce adversarial documents that are not easily detected by both humans and algorithms. 
Our experiments on the MS MARCO dataset show that IDEM can not only achieve a high level of attack performance, but also generate correct and fluent adversarial documents as evaluated by both automatic and human assessments. 
As future work, we plan to expand our research to include more advanced GLMs, such as GPT-3~\cite{DBLP:conf/nips/BrownMRSKDNSSAA20}.

\section*{Limitations}
In our experiments, as NRMs with cross-encoder are widely used, we focus on evaluating the textual adversarial robustness during the re-ranking stage and do not currently take into account the effect on the retrieval stage. 
But actually, in a ``first retrieval then re-ranking'' ranking paradigm, the attack is effective only when the adversarial documents are passed into the top retrieval results.
Meanwhile, dense retrieval (DR) models have been widely studied, and they may also inherit adversarial vulnerabilities due to the basics of PLMs. 
Besides, due to limitations in our computing resources, we only tested adding adversarial text to relatively short documents (i.e., passage-level), but the document content in real-world applications could be much longer. 
Therefore, further comprehensive investigations on examining the NRMs with different architectures, the effects of attacks on the retrieval models, and the manipulations on longer documents are left for future work.

\section*{Ethics Statement}
In this paper, we investigate the potential vulnerability concerns of the neural information retrieval (IR) systems and propose a document manipulation framework that generates adversarial documents that are not easily detected by both humans and IR systems.
We hope that this study could inspire further exploration and design of adversarial ranking defense/detection methods and aid in the development of robust real-world search engines.

\bibliography{main}

\begin{thebibliography}{45}
\expandafter\ifx\csname natexlab\endcsname\relax\def\natexlab#1{#1}\fi

\bibitem[{Brown et~al.(2020)Brown, Mann, Ryder, Subbiah, Kaplan, Dhariwal,
  Neelakantan, Shyam, Sastry, Askell, Agarwal, Herbert{-}Voss, Krueger,
  Henighan, Child, Ramesh, Ziegler, Wu, Winter, Hesse, Chen, Sigler, Litwin,
  Gray, Chess, Clark, Berner, McCandlish, Radford, Sutskever, and
  Amodei}]{DBLP:conf/nips/BrownMRSKDNSSAA20}
Tom~B. Brown, Benjamin Mann, Nick Ryder, Melanie Subbiah, Jared Kaplan,
  Prafulla Dhariwal, Arvind Neelakantan, Pranav Shyam, Girish Sastry, Amanda
  Askell, Sandhini Agarwal, Ariel Herbert{-}Voss, Gretchen Krueger, Tom
  Henighan, Rewon Child, Aditya Ramesh, Daniel~M. Ziegler, Jeffrey Wu, Clemens
  Winter, Christopher Hesse, Mark Chen, Eric Sigler, Mateusz Litwin, Scott
  Gray, Benjamin Chess, Jack Clark, Christopher Berner, Sam McCandlish, Alec
  Radford, Ilya Sutskever, and Dario Amodei. 2020.
\newblock \href
  {https://proceedings.neurips.cc/paper/2020/hash/1457c0d6bfcb4967418bfb8ac142f64a-Abstract.html}
  {Language models are few-shot learners}.
\newblock In \emph{Advances in Neural Information Processing Systems 33: Annual
  Conference on Neural Information Processing Systems 2020, NeurIPS 2020,
  December 6-12, 2020, virtual}.

\bibitem[{Castillo and Davison(2010)}]{DBLP:journals/ftir/CastilloD10}
Carlos Castillo and Brian~D. Davison. 2010.
\newblock \href {https://doi.org/10.1561/1500000021} {Adversarial web search}.
\newblock \emph{Found. Trends Inf. Retr.}, 4(5):377--486.

\bibitem[{Chen et~al.(2023)Chen, He, Hui, Sun, and
  Sun}]{DBLP:journals/ipm/ChenHHSS23}
Xuanang Chen, Ben He, Kai Hui, Le~Sun, and Yingfei Sun. 2023.
\newblock \href {https://doi.org/10.1016/j.ipm.2022.103135} {Dealing with
  textual noise for robust and effective {BERT} re-ranking}.
\newblock \emph{Inf. Process. Manag.}, 60(1):103135.

\bibitem[{Dalvi et~al.(2004)Dalvi, Domingos, Mausam, Sanghai, and
  Verma}]{DBLP:conf/kdd/DalviDMSV04}
Nilesh~N. Dalvi, Pedro~M. Domingos, Mausam, Sumit~K. Sanghai, and Deepak Verma.
  2004.
\newblock \href {https://doi.org/10.1145/1014052.1014066} {Adversarial
  classification}.
\newblock In \emph{Proceedings of the Tenth {ACM} {SIGKDD} International
  Conference on Knowledge Discovery and Data Mining, Seattle, Washington, USA,
  August 22-25, 2004}, pages 99--108. {ACM}.

\bibitem[{Davison et~al.(2006)Davison, Najork, and
  Converse}]{DBLP:journals/sigir/DavisonNC06}
Brian~D. Davison, Marc Najork, and Tim Converse. 2006.
\newblock \href {https://doi.org/10.1145/1189702.1189706} {Adversarial
  information retrieval on the web (airweb 2006)}.
\newblock \emph{{SIGIR} Forum}, 40(2):27--30.

\bibitem[{Devlin et~al.(2019)Devlin, Chang, Lee, and
  Toutanova}]{devlin-etal-2019-bert}
Jacob Devlin, Ming-Wei Chang, Kenton Lee, and Kristina Toutanova. 2019.
\newblock \href {https://doi.org/10.18653/v1/N19-1423} {{BERT}: Pre-training of
  deep bidirectional transformers for language understanding}.
\newblock In \emph{Proceedings of the 2019 Conference of the North {A}merican
  Chapter of the Association for Computational Linguistics: Human Language
  Technologies, Volume 1 (Long and Short Papers)}, pages 4171--4186,
  Minneapolis, Minnesota. Association for Computational Linguistics.

\bibitem[{Donahue et~al.(2020)Donahue, Lee, and
  Liang}]{donahue-etal-2020-enabling}
Chris Donahue, Mina Lee, and Percy Liang. 2020.
\newblock \href {https://doi.org/10.18653/v1/2020.acl-main.225} {Enabling
  language models to fill in the blanks}.
\newblock In \emph{Proceedings of the 58th Annual Meeting of the Association
  for Computational Linguistics}, pages 2492--2501, Online. Association for
  Computational Linguistics.

\bibitem[{Ebrahimi et~al.(2018)Ebrahimi, Rao, Lowd, and
  Dou}]{ebrahimi-etal-2018-hotflip}
Javid Ebrahimi, Anyi Rao, Daniel Lowd, and Dejing Dou. 2018.
\newblock \href {https://doi.org/10.18653/v1/P18-2006} {{H}ot{F}lip: White-box
  adversarial examples for text classification}.
\newblock In \emph{Proceedings of the 56th Annual Meeting of the Association
  for Computational Linguistics (Volume 2: Short Papers)}, pages 31--36,
  Melbourne, Australia. Association for Computational Linguistics.

\bibitem[{Fan et~al.(2018)Fan, Lewis, and Dauphin}]{fan-etal-2018-hierarchical}
Angela Fan, Mike Lewis, and Yann Dauphin. 2018.
\newblock \href {https://doi.org/10.18653/v1/P18-1082} {Hierarchical neural
  story generation}.
\newblock In \emph{Proceedings of the 56th Annual Meeting of the Association
  for Computational Linguistics (Volume 1: Long Papers)}, pages 889--898,
  Melbourne, Australia. Association for Computational Linguistics.

\bibitem[{Farooq(2019)}]{DBLP:journals/corr/abs-1911-11060}
Saad Farooq. 2019.
\newblock \href {http://arxiv.org/abs/1911.11060} {A survey on adversarial
  information retrieval on the web}.
\newblock \emph{CoRR}, abs/1911.11060.

\bibitem[{Goren et~al.(2018)Goren, Kurland, Tennenholtz, and
  Raiber}]{DBLP:conf/sigir/GorenKTR18}
Gregory Goren, Oren Kurland, Moshe Tennenholtz, and Fiana Raiber. 2018.
\newblock \href {https://doi.org/10.1145/3209978.3210012} {Ranking robustness
  under adversarial document manipulations}.
\newblock In \emph{The 41st International {ACM} {SIGIR} Conference on Research
  {\&} Development in Information Retrieval, {SIGIR} 2018, Ann Arbor, MI, USA,
  July 08-12, 2018}, pages 395--404. {ACM}.

\bibitem[{Goren et~al.(2020)Goren, Kurland, Tennenholtz, and
  Raiber}]{DBLP:conf/sigir/GorenKTR20}
Gregory Goren, Oren Kurland, Moshe Tennenholtz, and Fiana Raiber. 2020.
\newblock \href {https://doi.org/10.1145/3397271.3401058} {Ranking-incentivized
  quality preserving content modification}.
\newblock In \emph{Proceedings of the 43rd International {ACM} {SIGIR}
  conference on research and development in Information Retrieval, {SIGIR}
  2020, Virtual Event, China, July 25-30, 2020}, pages 259--268. {ACM}.

\bibitem[{Gy{\"{o}}ngyi and
  Garcia{-}Molina(2005)}]{DBLP:conf/airweb/GyongyiG05}
Zolt{\'{a}}n Gy{\"{o}}ngyi and Hector Garcia{-}Molina. 2005.
\newblock \href {http://airweb.cse.lehigh.edu/2005/gyongyi.pdf} {Web spam
  taxonomy}.
\newblock In \emph{AIRWeb 2005, First International Workshop on Adversarial
  Information Retrieval on the Web, co-located with the {WWW} conference,
  Chiba, Japan, May 2005}, pages 39--47.

\bibitem[{Hofst{\"{a}}tter et~al.(2021)Hofst{\"{a}}tter, Lipani, Althammer,
  Zlabinger, and Hanbury}]{DBLP:conf/ecir/HofstatterLAZH21}
Sebastian Hofst{\"{a}}tter, Aldo Lipani, Sophia Althammer, Markus Zlabinger,
  and Allan Hanbury. 2021.
\newblock \href {https://doi.org/10.1007/978-3-030-72113-8\_16} {Mitigating the
  position bias of transformer models in passage re-ranking}.
\newblock In \emph{Advances in Information Retrieval - 43rd European Conference
  on {IR} Research, {ECIR} 2021, Virtual Event, March 28 - April 1, 2021,
  Proceedings, Part {I}}, volume 12656 of \emph{Lecture Notes in Computer
  Science}, pages 238--253. Springer.

\bibitem[{Jiang et~al.(2021)Jiang, Tang, Xin, and Lin}]{jiang-etal-2021-bert}
Zhiying Jiang, Raphael Tang, Ji~Xin, and Jimmy Lin. 2021.
\newblock \href {https://doi.org/10.18653/v1/2021.blackboxnlp-1.39} {How does
  {BERT} rerank passages? an attribution analysis with information
  bottlenecks}.
\newblock In \emph{Proceedings of the Fourth BlackboxNLP Workshop on Analyzing
  and Interpreting Neural Networks for NLP}, pages 496--509, Punta Cana,
  Dominican Republic. Association for Computational Linguistics.

\bibitem[{Kann et~al.(2018)Kann, Rothe, and
  Filippova}]{kann-etal-2018-sentence}
Katharina Kann, Sascha Rothe, and Katja Filippova. 2018.
\newblock \href {https://doi.org/10.18653/v1/K18-1031} {Sentence-level fluency
  evaluation: References help, but can be spared!}
\newblock In \emph{Proceedings of the 22nd Conference on Computational Natural
  Language Learning}, pages 313--323, Brussels, Belgium. Association for
  Computational Linguistics.

\bibitem[{Karpukhin et~al.(2020)Karpukhin, Oguz, Min, Lewis, Wu, Edunov, Chen,
  and Yih}]{karpukhin-etal-2020-dense}
Vladimir Karpukhin, Barlas Oguz, Sewon Min, Patrick Lewis, Ledell Wu, Sergey
  Edunov, Danqi Chen, and Wen-tau Yih. 2020.
\newblock \href {https://doi.org/10.18653/v1/2020.emnlp-main.550} {Dense
  passage retrieval for open-domain question answering}.
\newblock In \emph{Proceedings of the 2020 Conference on Empirical Methods in
  Natural Language Processing (EMNLP)}, pages 6769--6781, Online. Association
  for Computational Linguistics.

\bibitem[{Kurland and Tennenholtz(2022)}]{DBLP:conf/sigir/KurlandT22}
Oren Kurland and Moshe Tennenholtz. 2022.
\newblock \href {https://doi.org/10.1145/3477495.3532771} {Competitive search}.
\newblock In \emph{{SIGIR} '22: The 45th International {ACM} {SIGIR} Conference
  on Research and Development in Information Retrieval, Madrid, Spain, July 11
  - 15, 2022}, pages 2838--2849. {ACM}.

\bibitem[{Kwiatkowski et~al.(2019)Kwiatkowski, Palomaki, Redfield, Collins,
  Parikh, Alberti, Epstein, Polosukhin, Devlin, Lee, Toutanova, Jones, Kelcey,
  Chang, Dai, Uszkoreit, Le, and Petrov}]{DBLP:journals/tacl/KwiatkowskiPRCP19}
Tom Kwiatkowski, Jennimaria Palomaki, Olivia Redfield, Michael Collins,
  Ankur~P. Parikh, Chris Alberti, Danielle Epstein, Illia Polosukhin, Jacob
  Devlin, Kenton Lee, Kristina Toutanova, Llion Jones, Matthew Kelcey,
  Ming{-}Wei Chang, Andrew~M. Dai, Jakob Uszkoreit, Quoc Le, and Slav Petrov.
  2019.
\newblock \href {https://doi.org/10.1162/tacl\_a\_00276} {Natural questions: a
  benchmark for question answering research}.
\newblock \emph{Trans. Assoc. Comput. Linguistics}, 7:452--466.

\bibitem[{Lei et~al.(2022)Lei, Cao, Li, Zhou, Fang, and
  Pechenizkiy}]{lei-etal-2022-phrase}
Yibin Lei, Yu~Cao, Dianqi Li, Tianyi Zhou, Meng Fang, and Mykola Pechenizkiy.
  2022.
\newblock \href {https://doi.org/10.18653/v1/2022.findings-naacl.83}
  {Phrase-level textual adversarial attack with label preservation}.
\newblock In \emph{Findings of the Association for Computational Linguistics:
  NAACL 2022}, pages 1095--1112, Seattle, United States. Association for
  Computational Linguistics.

\bibitem[{Leng et~al.(2012)Leng, Patchmuthu, Singh, and
  Mohan}]{DBLP:journals/cas/LengKSM12}
Alex Goh~Kwang Leng, Ravi~Kumar Patchmuthu, Ashutosh~Kumar Singh, and Anand
  Mohan. 2012.
\newblock \href {https://doi.org/10.1080/01969722.2012.707491} {Link-based spam
  algorithms in adversarial information retrieval}.
\newblock \emph{Cybern. Syst.}, 43(6):459--475.

\bibitem[{Lewis et~al.(2020)Lewis, Liu, Goyal, Ghazvininejad, Mohamed, Levy,
  Stoyanov, and Zettlemoyer}]{lewis-etal-2020-bart}
Mike Lewis, Yinhan Liu, Naman Goyal, Marjan Ghazvininejad, Abdelrahman Mohamed,
  Omer Levy, Veselin Stoyanov, and Luke Zettlemoyer. 2020.
\newblock \href {https://doi.org/10.18653/v1/2020.acl-main.703} {{BART}:
  Denoising sequence-to-sequence pre-training for natural language generation,
  translation, and comprehension}.
\newblock In \emph{Proceedings of the 58th Annual Meeting of the Association
  for Computational Linguistics}, pages 7871--7880, Online. Association for
  Computational Linguistics.

\bibitem[{Lin et~al.(2021)Lin, Nogueira, and
  Yates}]{DBLP:series/synthesis/2021LinNY}
Jimmy Lin, Rodrigo Nogueira, and Andrew Yates. 2021.
\newblock \href {https://doi.org/10.2200/S01123ED1V01Y202108HLT053}
  {\emph{Pretrained Transformers for Text Ranking: {BERT} and Beyond}}.
\newblock Synthesis Lectures on Human Language Technologies. Morgan {\&}
  Claypool Publishers.

\bibitem[{Liu et~al.(2022)Liu, Kang, Tang, Song, Sun, Wang, Lu, and
  Liu}]{DBLP:conf/ccs/LiuKTSSWLL22}
Jiawei Liu, Yangyang Kang, Di~Tang, Kaisong Song, Changlong Sun, Xiaofeng Wang,
  Wei Lu, and Xiaozhong Liu. 2022.
\newblock \href {https://doi.org/10.1145/3548606.3560683} {Order-disorder:
  Imitation adversarial attacks for black-box neural ranking models}.
\newblock In \emph{Proceedings of the 2022 {ACM} {SIGSAC} Conference on
  Computer and Communications Security, {CCS} 2022, Los Angeles, CA, USA,
  November 7-11, 2022}, pages 2025--2039. {ACM}.

\bibitem[{Liu et~al.(2021)Liu, Yuan, Fu, Jiang, Hayashi, and
  Neubig}]{DBLP:journals/corr/abs-2107-13586}
Pengfei Liu, Weizhe Yuan, Jinlan Fu, Zhengbao Jiang, Hiroaki Hayashi, and
  Graham Neubig. 2021.
\newblock \href {http://arxiv.org/abs/2107.13586} {Pre-train, prompt, and
  predict: {A} systematic survey of prompting methods in natural language
  processing}.
\newblock \emph{CoRR}, abs/2107.13586.

\bibitem[{Madry et~al.(2018)Madry, Makelov, Schmidt, Tsipras, and
  Vladu}]{DBLP:conf/iclr/MadryMSTV18}
Aleksander Madry, Aleksandar Makelov, Ludwig Schmidt, Dimitris Tsipras, and
  Adrian Vladu. 2018.
\newblock \href {https://openreview.net/forum?id=rJzIBfZAb} {Towards deep
  learning models resistant to adversarial attacks}.
\newblock In \emph{6th International Conference on Learning Representations,
  {ICLR} 2018, Vancouver, BC, Canada, April 30 - May 3, 2018, Conference Track
  Proceedings}. OpenReview.net.

\bibitem[{Nguyen et~al.(2016)Nguyen, Rosenberg, Song, Gao, Tiwary, Majumder,
  and Deng}]{DBLP:conf/nips/NguyenRSGTMD16}
Tri Nguyen, Mir Rosenberg, Xia Song, Jianfeng Gao, Saurabh Tiwary, Rangan
  Majumder, and Li~Deng. 2016.
\newblock \href {http://ceur-ws.org/Vol-1773/CoCoNIPS\_2016\_paper9.pdf} {{MS}
  {MARCO:} {A} human generated machine reading comprehension dataset}.
\newblock In \emph{Proceedings of the Workshop on Cognitive Computation:
  Integrating neural and symbolic approaches 2016 co-located with the 30th
  Annual Conference on Neural Information Processing Systems {(NIPS} 2016),
  Barcelona, Spain, December 9, 2016}, volume 1773 of \emph{{CEUR} Workshop
  Proceedings}. CEUR-WS.org.

\bibitem[{Nogueira and Cho(2019)}]{DBLP:journals/corr/abs-1901-04085}
Rodrigo~Frassetto Nogueira and Kyunghyun Cho. 2019.
\newblock \href {http://arxiv.org/abs/1901.04085} {Passage re-ranking with
  {BERT}}.
\newblock \emph{CoRR}, abs/1901.04085.

\bibitem[{Ntoulas et~al.(2006)Ntoulas, Najork, Manasse, and
  Fetterly}]{DBLP:conf/www/NtoulasNMF06}
Alexandros Ntoulas, Marc Najork, Mark~S. Manasse, and Dennis Fetterly. 2006.
\newblock \href {https://doi.org/10.1145/1135777.1135794} {Detecting spam web
  pages through content analysis}.
\newblock In \emph{Proceedings of the 15th international conference on World
  Wide Web, {WWW} 2006, Edinburgh, Scotland, UK, May 23-26, 2006}, pages
  83--92. {ACM}.

\bibitem[{Penha et~al.(2022)Penha, C{\^{a}}mara, and
  Hauff}]{DBLP:conf/ecir/PenhaCH22}
Gustavo Penha, Arthur C{\^{a}}mara, and Claudia Hauff. 2022.
\newblock \href {https://doi.org/10.1007/978-3-030-99736-6\_27} {Evaluating the
  robustness of retrieval pipelines with query variation generators}.
\newblock In \emph{Advances in Information Retrieval - 44th European Conference
  on {IR} Research, {ECIR} 2022, Stavanger, Norway, April 10-14, 2022,
  Proceedings, Part {I}}, volume 13185 of \emph{Lecture Notes in Computer
  Science}, pages 397--412. Springer.

\bibitem[{Radford et~al.(2019)Radford, Wu, Child, Luan, Amodei, Sutskever
  et~al.}]{radford2019language}
Alec Radford, Jeffrey Wu, Rewon Child, David Luan, Dario Amodei, Ilya
  Sutskever, et~al. 2019.
\newblock \href
  {https://d4mucfpksywv.cloudfront.net/better-language-models/language-models.pdf}
  {Language models are unsupervised multitask learners}.
\newblock \emph{OpenAI blog}, 1(8):9.

\bibitem[{Raffel et~al.(2020)Raffel, Shazeer, Roberts, Lee, Narang, Matena,
  Zhou, Li, and Liu}]{DBLP:journals/jmlr/RaffelSRLNMZLL20}
Colin Raffel, Noam Shazeer, Adam Roberts, Katherine Lee, Sharan Narang, Michael
  Matena, Yanqi Zhou, Wei Li, and Peter~J. Liu. 2020.
\newblock \href {http://jmlr.org/papers/v21/20-074.html} {Exploring the limits
  of transfer learning with a unified text-to-text transformer}.
\newblock \emph{J. Mach. Learn. Res.}, 21:140:1--140:67.

\bibitem[{Raifer et~al.(2017)Raifer, Raiber, Tennenholtz, and
  Kurland}]{DBLP:conf/sigir/RaiferRTK17}
Nimrod Raifer, Fiana Raiber, Moshe Tennenholtz, and Oren Kurland. 2017.
\newblock \href {https://doi.org/10.1145/3077136.3080785} {Information
  retrieval meets game theory: The ranking competition between documents?
  authors}.
\newblock In \emph{Proceedings of the 40th International {ACM} {SIGIR}
  Conference on Research and Development in Information Retrieval, Shinjuku,
  Tokyo, Japan, August 7-11, 2017}, pages 465--474. {ACM}.

\bibitem[{Raval and Verma(2020)}]{DBLP:journals/corr/abs-2008-02197}
Nisarg Raval and Manisha Verma. 2020.
\newblock \href {http://arxiv.org/abs/2008.02197} {One word at a time:
  adversarial attacks on retrieval models}.
\newblock \emph{CoRR}, abs/2008.02197.

\bibitem[{Reimers and Gurevych(2019)}]{reimers-gurevych-2019-sentence}
Nils Reimers and Iryna Gurevych. 2019.
\newblock \href {https://doi.org/10.18653/v1/D19-1410} {Sentence-{BERT}:
  Sentence embeddings using {S}iamese {BERT}-networks}.
\newblock In \emph{Proceedings of the 2019 Conference on Empirical Methods in
  Natural Language Processing and the 9th International Joint Conference on
  Natural Language Processing (EMNLP-IJCNLP)}, pages 3982--3992, Hong Kong,
  China. Association for Computational Linguistics.

\bibitem[{Robertson et~al.(1995)Robertson, Walker, Hancock{-}Beaulieu, Gatford,
  and Payne}]{DBLP:conf/trec/RobertsonWHGP95}
Stephen~E. Robertson, Steve Walker, Micheline Hancock{-}Beaulieu, Mike Gatford,
  and A.~Payne. 1995.
\newblock \href {http://trec.nist.gov/pubs/trec4/papers/city.ps.gz} {Okapi at
  {TREC-4}}.
\newblock In \emph{Proceedings of The Fourth Text REtrieval Conference, {TREC}
  1995, Gaithersburg, Maryland, USA, November 1-3, 1995}, volume 500-236 of
  \emph{{NIST} Special Publication}. National Institute of Standards and
  Technology {(NIST)}.

\bibitem[{Song et~al.(2020)Song, Rush, and
  Shmatikov}]{song-etal-2020-adversarial}
Congzheng Song, Alexander Rush, and Vitaly Shmatikov. 2020.
\newblock \href {https://doi.org/10.18653/v1/2020.emnlp-main.344} {Adversarial
  semantic collisions}.
\newblock In \emph{Proceedings of the 2020 Conference on Empirical Methods in
  Natural Language Processing (EMNLP)}, pages 4198--4210, Online. Association
  for Computational Linguistics.

\bibitem[{Song et~al.(2022)Song, Zhang, Zhu, Tang, and
  Yang}]{song-etal-2022-trattack}
Junshuai Song, Jiangshan Zhang, Jifeng Zhu, Mengyun Tang, and Yong Yang. 2022.
\newblock \href {https://doi.org/10.18653/v1/2022.repl4nlp-1.20} {{TRA}ttack:
  Text rewriting attack against text retrieval}.
\newblock In \emph{Proceedings of the 7th Workshop on Representation Learning
  for NLP}, pages 191--203, Dublin, Ireland. Association for Computational
  Linguistics.

\bibitem[{Wang et~al.(2020)Wang, Wei, Dong, Bao, Yang, and
  Zhou}]{DBLP:conf/nips/WangW0B0020}
Wenhui Wang, Furu Wei, Li~Dong, Hangbo Bao, Nan Yang, and Ming Zhou. 2020.
\newblock \href
  {https://proceedings.neurips.cc/paper/2020/hash/3f5ee243547dee91fbd053c1c4a845aa-Abstract.html}
  {Minilm: Deep self-attention distillation for task-agnostic compression of
  pre-trained transformers}.
\newblock In \emph{Advances in Neural Information Processing Systems 33: Annual
  Conference on Neural Information Processing Systems 2020, NeurIPS 2020,
  December 6-12, 2020, virtual}.

\bibitem[{Wang et~al.(2022)Wang, Lyu, and Anand}]{DBLP:conf/ictir/WangLA22}
Yumeng Wang, Lijun Lyu, and Avishek Anand. 2022.
\newblock \href {https://doi.org/10.1145/3539813.3545122} {{BERT} rankers are
  brittle: {A} study using adversarial document perturbations}.
\newblock In \emph{{ICTIR} '22: The 2022 {ACM} {SIGIR} International Conference
  on the Theory of Information Retrieval, Madrid, Spain, July 11 - 12, 2022},
  pages 115--120. {ACM}.

\bibitem[{Webber et~al.(2010)Webber, Moffat, and
  Zobel}]{DBLP:journals/tois/WebberMZ10}
William Webber, Alistair Moffat, and Justin Zobel. 2010.
\newblock \href {https://doi.org/10.1145/1852102.1852106} {A similarity measure
  for indefinite rankings}.
\newblock \emph{{ACM} Trans. Inf. Syst.}, 28(4):20:1--20:38.

\bibitem[{Wu et~al.(2022)Wu, Zhang, Guo, de~Rijke, Fan, and
  Cheng}]{DBLP:journals/corr/abs-2204-01321}
Chen Wu, Ruqing Zhang, Jiafeng Guo, Maarten de~Rijke, Yixing Fan, and Xueqi
  Cheng. 2022.
\newblock \href {https://doi.org/10.48550/arXiv.2204.01321} {{PRADA:} practical
  black-box adversarial attacks against neural ranking models}.
\newblock \emph{CoRR}, abs/2204.01321.

\bibitem[{Xu and Du(2020)}]{DBLP:journals/eaai/XuD20}
Jincheng Xu and Qingfeng Du. 2020.
\newblock \href {https://doi.org/10.1016/j.engappai.2020.103641} {Texttricker:
  Loss-based and gradient-based adversarial attacks on text classification
  models}.
\newblock \emph{Eng. Appl. Artif. Intell.}, 92:103641.

\bibitem[{Zhuang and Zuccon(2021)}]{zhuang-zuccon-2021-dealing}
Shengyao Zhuang and Guido Zuccon. 2021.
\newblock \href {https://doi.org/10.18653/v1/2021.emnlp-main.225} {Dealing with
  typos for {BERT}-based passage retrieval and ranking}.
\newblock In \emph{Proceedings of the 2021 Conference on Empirical Methods in
  Natural Language Processing}, pages 2836--2842, Online and Punta Cana,
  Dominican Republic. Association for Computational Linguistics.

\bibitem[{Zou et~al.(2021)Zou, Zhang, Cai, Ma, Cheng, Wang, Shi, Cheng, and
  Yin}]{DBLP:conf/kdd/ZouZCMCWSCY21}
Lixin Zou, Shengqiang Zhang, Hengyi Cai, Dehong Ma, Suqi Cheng, Shuaiqiang
  Wang, Daiting Shi, Zhicong Cheng, and Dawei Yin. 2021.
\newblock \href {https://doi.org/10.1145/3447548.3467147} {Pre-trained language
  model based ranking in baidu search}.
\newblock In \emph{{KDD} '21: The 27th {ACM} {SIGKDD} Conference on Knowledge
  Discovery and Data Mining, Virtual Event, Singapore, August 14-18, 2021},
  pages 4014--4022. {ACM}.

\end{thebibliography}
\bibliographystyle{acl_natbib}

\clearpage
\appendix

\section{Appendix}
\subsection{Details of Surrogate NRMs}\label{sec.app.surro}
As mentioned in Section~\ref{sec.exp.setup}, we conduct black-box attacks against the victim neural ranking model (NRM) $\mathcal{M}_{V}$ using three types of surrogate NRMs. 
Since the training data of the victim NRM is typically unavailable, we use Eval queries in the MS MARCO dataset rather than train queries to construct the surrogate training data $\mathcal{T}_{S}$. 
As we move from $\mathcal{M}_{S_1}$ to $\mathcal{M}_{S_2}$, the number of in-domain Eval queries used for ranking imitation decreases from 6,837 to 200, it means the frequency of querying the victim NRM is greatly reduced. 
For each Eval query $q_{i}$, we collect all 406 document pairs $(d_{i,j}, d_{i,k})$ ($1\!\leq\!j\!<\!k\!\leq\!29$) from the top-29 of the re-ranking list produced by the victim NRM over the top-1K BM25 candidates, and use this to construct the surrogate training data $\mathcal{T}_{S}$ as described in Section~\ref{sec.problem}.
As a result, the in-domain surrogate training data for $\mathcal{M}_{S_1}$ and $\mathcal{M}_{S_2}$ contains 81.2 thousand and 2.77 million training triples, respectively.
Additionally, $\mathcal{M}_{S_3}$ implies the scenario where no in-domain data from the victim NRM is available, and we collect 3.72 million training triples from the NQ dataset as the out-of-domain (OOD) surrogate training data.
The surrogate NRMs are based on the pre-trained BERT-Base model and fine-tuned for two epochs with a learning rate of 3e-6 and batch size of 16 for $\mathcal{M}_{S_1}$ and $\mathcal{M}_{S_2}$, and one epoch for $\mathcal{M}_{S_3}$ with the same learning rate and batch size. 
It is important to note that the goal of this work is not to develop a new training method for the surrogate NRMs, and thus we directly adopt the hinge loss with a margin of 1 for ranking imitation as per previous work~\cite{DBLP:journals/corr/abs-2204-01321}.

\subsection{Details of Attack Baselines}\label{sec.app.baseline}
In our adversarial attack experiments, we examine the following baseline methods:

\textbf{Query+}~\cite{DBLP:conf/ccs/LiuKTSSWLL22} is an intuitive baseline that directly appends the query text to the beginning of the target document.
Although the query text can be placed at any position in the target document, or even determined by our proposed position-aware mechanism, appending to the beginning usually produces greater attack results due to the positional bias in Transformer-based NRMs~\cite{jiang-etal-2021-bert,DBLP:conf/ecir/HofstatterLAZH21}. 
Thus, Query+ acts as a baseline method that does not take into account the invisibility aspect of the attack.

\textbf{PRADA}~\cite{DBLP:journals/corr/abs-2204-01321} is a Word Substitution Ranking Attack (WSRA) method, it first finds important words (i.e., sub-word tokens) in the target document according to the gradient magnitude~\cite{DBLP:journals/eaai/XuD20}, and then greedily replaces them with the synonyms found in a perturbed word embedding space via PGD~\cite{DBLP:conf/iclr/MadryMSTV18}.
Based on our observations, when attacking different random target samples, PRADA is able to attain attack performance (ASR) that is close to the results reported in its original publication. 
This is particularly true when the victim NRM has relatively poor ranking performance, due to the obvious fact that it is much easier to attack a weaker victim NRM.
Despite this variability, the overall conclusions drawn from our experimentation remain unchanged.

\textbf{Brittle-BERT}~\cite{DBLP:conf/ictir/WangLA22} studies both local (i.e., a particular query-document instance) and global (i.e., an entire workload of queries) ranking attack to cause a large rank demotion or promotion by adding/replacing a small number of tokens.
In our work, we only adopt the local setting and add tokens to the beginning of the target document as it usually produces better attack results. Specifically, Brittle-BERT first initializes a few placeholder tokens at the beginning of the target document and then employs HotFlip~\cite{ebrahimi-etal-2018-hotflip} algorithm to update them as being more adversarial.

\textbf{PAT}~\cite{DBLP:conf/ccs/LiuKTSSWLL22} generates and adds several trigger tokens at the beginning of the target document.
In addition to the ranking-incentivized objective, the search objective of PAT is equipped with semantic and fluency constraints using the pre-trained BERT model. 
The surrogate NRMs trained with hinge loss have a one-class prediction layer, but PAT needs a surrogate NRM with a two-class prediction layer, namely, `Pairwise BERT' as denoted in PAT~\cite{DBLP:conf/ccs/LiuKTSSWLL22}, so we use the same surrogate training data to obtain `Pairwise BERT' using the default imitation loss in PAT.
 
In order to evaluate these baselines, we utilize their publicly available implementations and ensure that all settings are consistent with those described in their respective official publications.

\subsection{Time Cost of Attack}\label{sec.app.time}
\begin{figure}[t]
\centering
\includegraphics[width=0.98\linewidth]{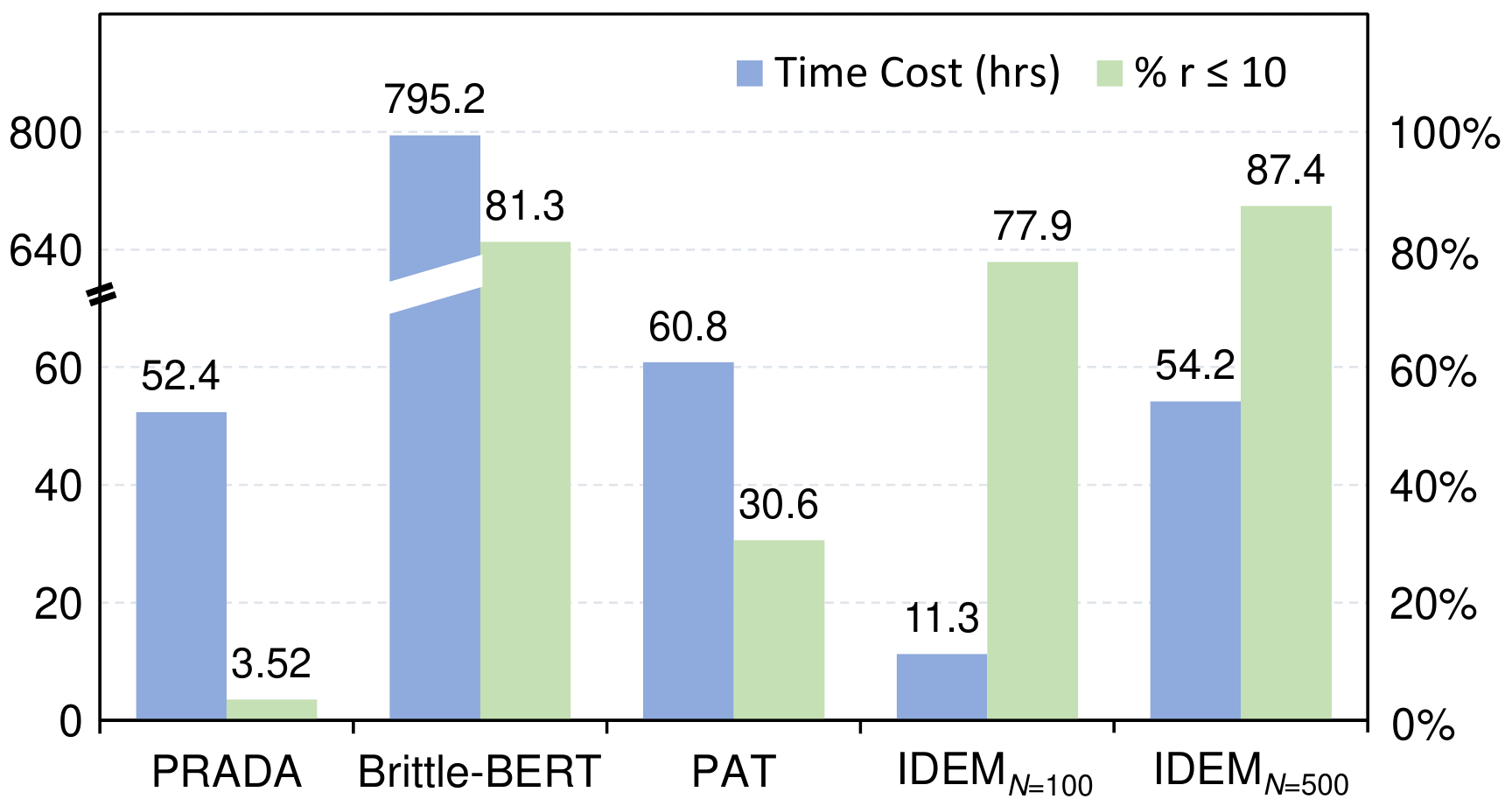}
\caption{The total time cost of generating 5K adversarial documents for different attack methods, along with the attack performance ($\% \, r\leq10$) on Easy-5 target documents under surrogate $\mathcal{M}_{S_{1}}$ (in Table~\ref{tab:attack.res}).}
\label{fig.time.cost}
\end{figure}

In previous PRADA, Brittle-BERT and PAT, the replacement, selection, and search of tokens are carried out one by one using the surrogate NRM to produce an adversarial document, so it needs a large amount of time to complete the attack process.
However, in our proposed IDEM framework, the adversarial text is first generated via a GLM (e.g., BART), and then combined in the sentence level, which is more efficient than the token level.
As seen in Figure~\ref{fig.time.cost}, we summarize the total time cost of different attack methods in producing 5,000 adversarial documents using one Titan RTX GPU. 
When at most 100 candidate connection sentences are generated for the position-wise combination, IDEM$_{N=100}$ only takes 11.3 hours to achieve great attack performance.
By comparison, PRADA and PAT consume more time but still perform worse in attack, Brittle-BERT takes a huge time cost (near 800 hours) even though its attack performance is considerable.
Additionally, compared with Brittle-BERT, IDEM$_{N=500}$ can produce comparable attack performance but take much less time cost.
Therefore, our proposed IDEM method lays equal stress on attack efficiency and performance.

\begin{figure}
\centering
\includegraphics[width=0.98\columnwidth]{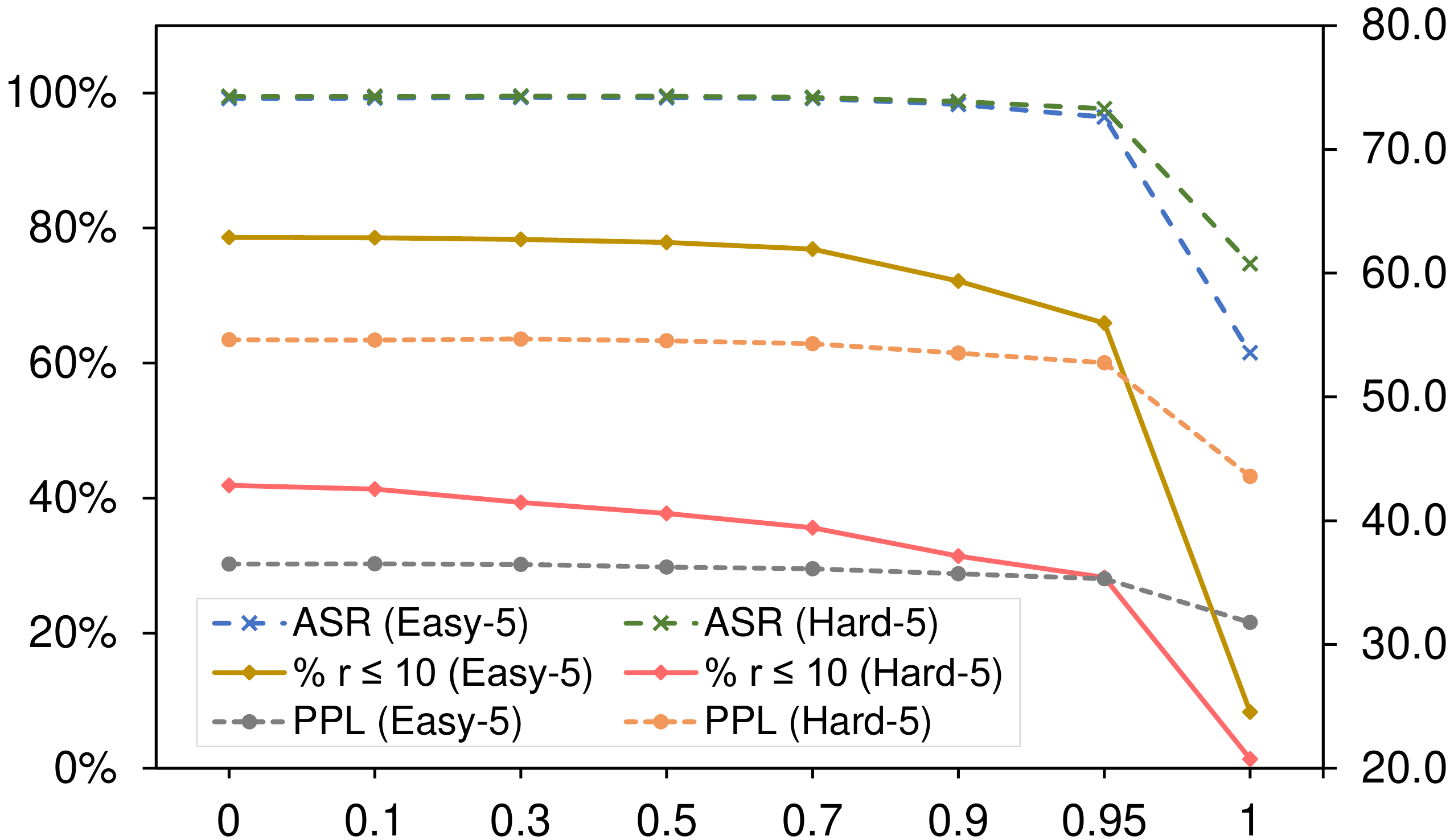}
\caption{The impact of $\alpha$ in the merging score (Eq.~\ref{eq.final}) on the attack results of IDEM$_{N=100}$.}
\label{fig.aplha}
\end{figure}
\begin{table*}[t]
\centering
\resizebox{\textwidth}{!}{
\begin{tabular}{l|l|cc}
\toprule
  \textbf{Method} & \textbf{Original or Adversarial Text} & \textbf{Rank}$\downarrow$ & \textbf{PPL}$\downarrow$ \\
  \midrule
  \multirow{4}{*}{Original} & Carry your baby in a sling or front carrier. Feeling your baby's warmth, smelling & \multirow{4}{*}{88} & \multirow{4}{*}{33.9} \\ 
  & his sweet scent, and looking down often to make eye contact with him can help & \\
  & you bond. Spend plenty of close-up face time with your baby. Smile at him, and  & \\ 
  & return the smile when he smiles first. & \\
  \midrule
  \multirow{2}{*}{Query+} & \textcolor{Red}{what age can you wear baby on back in a carrier?} Carry your baby in a sling or fr- & \multirow{2}{*}{2} & \multirow{2}{*}{40.4} \\
  & ont carrier. Feeling your baby's warmth, smelling his sweet scent, and looking ... & \\
  \midrule
  \multicolumn{4}{c}{Surrogate NRM $\mathcal{M}_{S_1}$} \\
  \midrule
  \multirow{3}{*}{PRADA} & \textcolor{Red}{wear} your baby in a sling \textcolor{Red}{ord} front carrier. Feeling your baby's \textcolor{Red}{warm}, smelling his & \multirow{3}{*}{27} & \multirow{3}{*}{100.8} \\
  & sweet scent, and looking down often to make eye contact with him can help you \textcolor{Red}{b-} & \\
  & \textcolor{Red}{ondage}. Spend plenty of close-up face time with your baby. Smile at him, and ... & \\
  \midrule
  \multirow{2}{*}{Brittle-BERT} & \textcolor{Red}{pendingerabidheartedivating aged aged 292,oning worn wear} Carry your baby in a & \multirow{2}{*}{2} & \multirow{2}{*}{125.0} \\
  & sling or front carrier. Feeling your baby's warmth, smelling his sweet scent, and ... & \\ 
  \midrule
  \multirow{2}{*}{PAT} & \textcolor{Red}{about 30 year old babies can carry baby carriers back to age} Carry your baby in a & \multirow{2}{*}{2} & \multirow{2}{*}{52.0} \\
  & sling or front carrier. Feeling your baby's warmth, smelling his sweet scent, and ... & \\
  \midrule
  \multirow{2}{*}{IDEM$_{N = 100}$} & Carry your baby in a sling or front carrier. \textcolor{Red}{A child of any age can wear shoes with} & \multirow{2}{*}{9} & \multirow{2}{*}{33.7} \\ 
  & \textcolor{Red}{a sling.} Feeling your baby's warmth, smelling his sweet scent, and looking ... & \\
  \midrule
  \multirow{2}{*}{IDEM$_{N = 500}$} & Carry your baby in a sling or front carrier. \textcolor{Red}{Most parents wear infant carriers arou-} & \multirow{2}{*}{1} & \multirow{2}{*}{36.5} \\ 
  & \textcolor{Red}{nd age 3, 4, and 5.} Feeling your baby's warmth, smelling his sweet scent, and ... & \\
  \midrule
  
  \multicolumn{4}{c}{Surrogate NRM $\mathcal{M}_{S_2}$} \\
  \midrule
  \multirow{2}{*}{PRADA} & Carry your baby in a \textcolor{Red}{slingshot ord} front carrier. Feeling your baby's \textcolor{Red}{warm}, smell- & \multirow{2}{*}{65} & \multirow{2}{*}{82.2} \\
  & ing his sweet scent, and looking down often to make eye contact with him can ... & \\
  \midrule
  \multirow{2}{*}{Brittle-BERT} & \textcolor{Red}{modernism age hms× chestnut beyonce rappers commercially whilst wearing md r-} & \multirow{2}{*}{14} & \multirow{2}{*}{175.2} \\
  & \textcolor{Red}{espectively} Carry your baby in a sling or front  carrier. Feeling your baby's ... & \\ 
  \midrule
  \multirow{2}{*}{PAT} & \textcolor{Red}{be a year old and can wear around} Carry your baby in a sling or front carrier. Feel- & \multirow{2}{*}{2} & \multirow{2}{*}{48.3} \\
  & ing your baby's warmth, smelling his sweet scent, and looking down often to ... & \\
  \midrule
  \multirow{2}{*}{IDEM$_{N = 100}$} & Carry your baby in a sling or front carrier. \textcolor{Red}{Wearing your baby on back is always a} & \multirow{2}{*}{9} & \multirow{2}{*}{29.4} \\ 
  & \textcolor{Red}{good idea.} Feeling your baby's warmth, smelling his sweet scent, and looking ... & \\
  \midrule
  \multirow{2}{*}{IDEM$_{N = 500}$} & Carry your baby in a sling or front carrier. \textcolor{Red}{You can and should be wearing baby on } & \multirow{2}{*}{9} & \multirow{2}{*}{36.9} \\ 
  & \textcolor{Red}{back in a carrier.} Feeling your baby's warmth, smelling his sweet scent, and ... & \\
  \midrule

  \multicolumn{4}{c}{Surrogate NRM $\mathcal{M}_{S_3}$} \\
  \midrule
  \multirow{2}{*}{PRADA} & \textcolor{Red}{wear} your baby in a \textcolor{Red}{slingshot ord} front carrier. Feeling your baby's warmth, smell- & \multirow{2}{*}{28} & \multirow{2}{*}{67.2} \\
  & ing his sweet scent, and looking down often to make eye contact with him can ... & \\
  \midrule
  \multirow{2}{*}{Brittle-BERT} & \textcolor{Red}{offspring coherent examples declined toys widespread adulthood noun whether bu-} & \multirow{2}{*}{58} & \multirow{2}{*}{143.8} \\
  & \textcolor{Red}{ckled off wear} Carry your baby in a sling or front carrier. Feeling your baby's ... & \\ 
  \midrule
  \multirow{2}{*}{PAT} & \textcolor{Red}{for example, may carry twenty cents} Carry your baby in a sling or front carrier. Fe- & \multirow{2}{*}{107} & \multirow{2}{*}{49.4} \\
  & eling your baby's warmth, smelling his sweet scent, and looking down often to ... & \\
  \midrule
  \multirow{2}{*}{IDEM$_{N = 100}$} & \textcolor{Red}{A child of any age can wear shoes with a sling.} Carry your baby in a sling or front & \multirow{2}{*}{9} & \multirow{2}{*}{39.1} \\ 
  & carrier. Feeling your baby's warmth, smelling his sweet scent, and looking down ... & \\
  \midrule
  \multirow{2}{*}{IDEM$_{N = 500}$} & Carry your baby in a sling or front carrier. \textcolor{Red}{You can and should be wearing baby on} & \multirow{2}{*}{9} & \multirow{2}{*}{36.9} \\ 
  & \textcolor{Red}{back in a carrier.} Feeling your baby's warmth, smelling his sweet scent, and ... & \\

\bottomrule
\end{tabular}
}
\caption{Adversarial documents generated by various attack methods under three kinds of surrogate NRMs on the same related but irreverent document for the query ``what age can you wear baby on back in a carrier?'' from the MS MARCO Dev set. The inserted and perturbed words are marked as \textcolor{Red}{Red} for easy comparisons.}
\label{tab:cases.res}
\end{table*}

\subsection{The Impact of $\alpha$ in the Merging Score}\label{sec.app.alpha}
In our IDEM framework, after generating a series of connection sentences between the query and the target document, a position-aware merging mechanism is employed to decide the final adversarial document, wherein a coherence score and a relevance score are added together using $\alpha$ as seen in Eq.~\ref{eq.final}. 
As shown in Figure~\ref{fig.aplha}, we can observe that $\alpha$ in a wide range (from 0 to 0.95) does not affect the attack success rate (ASR) too much on both Easy-5 and Hard-5 target document sets, and the $\% \, r\leq10$ metric starts to decrease at $\alpha\!=\!0.5$ and $\alpha\!=\!0.1$ on Easy-5 and Hard-5 target document sets, respectively. 
As for the perplexity (PPL) metric (smaller is better), when $\alpha$ increases (more attention on the coherence), PPL value does not change a lot until $\alpha$ reaches about 0.9 to 1.
Meanwhile, it can produce adversarial documents with lower PPL than original ones, e.g., when $\alpha$ is 1, the average PPL points on Easy-5 and Hard-5 target document sets are only 31.8 and 43.6, respectively.

\subsection{Adversarial Examples}\label{sec.app.examples}
To better qualitatively understand how different attack methods work, we show adversarial examples produced by them under three types of surrogate NRMs in Table~\ref{tab:cases.res}.
We can observe that Query+, and Brittle-BERT, PAT, IDEM under $\mathcal{M}_{S_1}$ promote the target document ranked at 88th into top-10.
However, adding query text (i.e., Query+) and unnatural token sequence (i.e., Brittle-BERT and PAT) at the beginning make adversarial documents distinguishable, while the inserted adversarial text by IDEM is more semantically consistent with the original surrounding content.
When the surrogate NRM degrades from $\mathcal{M}_{S_1}$ to $\mathcal{M}_{S_2}$ as not enough in-domain training samples are available, we can see that the ranking of the adversarial document by PRADA decreases from 27th to 65th, and the ranking of the adversarial document by Brittle-BERT also decreases from 2nd to 14th.
Furthermore, when an OOD surrogate NRM (i.e., $\mathcal{M}_{S_3}$) is used due to the forbidden access to the victim NRM, we can find out that the attack effects of PRADA, Brittle-BERT and PAT are greatly suppressed.
For example, although PAT under $\mathcal{M}_{S_2}$ promotes the target document to the ranking of 2nd, PAT under $\mathcal{M}_{S_3}$ even demotes the ranking of the target document from 88th to 107th.
In contrast, under both $\mathcal{M}_{S_2}$ and $\mathcal{M}_{S_3}$, IDEM robustly promotes the target document into top-10 (i.e., 9th), and the fluency and correctness of adversarial documents are still within an acceptable range.
From these adversarial cases, it is evident that IDEM is less dependent on the surrogate NRM and can perform attacks more robustly than previous attack methods, indicating a flexible use condition of IDEM in real-world situations.

\end{document}